\RequirePackage{fix-cm}
\documentclass{svjour3}                     
\smartqed  
\usepackage{graphicx}
\usepackage{url}
\usepackage{natbib}
\usepackage{ulem}
\usepackage{amsmath}
\usepackage{color}
\usepackage{hyperref}
\newcommand{\arcsec}{\ensuremath{^{\prime\prime}}}
\newcommand{\od}[1]{\mbox{\textsc{od}\,\texttt{#1}}}
\newcommand{\obsid}[1]{\mbox{\textsc{obsid}\,\texttt{{#1}}}}
\begin{document}
%
%
%
\title{Relative pointing offset analysis of calibration targets with repeated observations with \textit{Herschel}-SPIRE Fourier-Transform Spectrometer \thanks{\textit{Herschel} is an ESA space observatory with science instruments provided by European-led Principal Investigator consortia and with important participation from NASA.}
}

\titlerunning{FTS relative pointing analysis}        

\author{Ivan Valtchanov         \and
        Rosalind Hopwood \and
        Edward Polehampton \and
        Dominique Benielli \and
        Trevor Fulton \and
        Peter Imhof \and
        Tomasz Konopczy\'nski \and 
        Tanya Lim \and
        Nanyao Lu \and
        Nicola Marchili \and
        David Naylor \and
        Bruce Swinyard
}

\authorrunning{Valtchanov et al.} 

\institute{I. Valtchanov \at
              Herschel Science Centre, ESAC, P.O. Box 78, 28691 Villanueva de la Ca\~nada, Madrid, Spain\\
              \email{ivaltchanov@sciops.esa.int}           
           \and
           R. Hopwood \at
           Physics Department, Imperial College London, South Kensington Campus, SW7 2AZ, UK
           \and
           E. Polehampton \at
           RAL Space, Rutherford Appleton Laboratory, Didcot OX11 0QX, UK;\\
           Institute for Space Imaging Science, Department of Physics \& Astronomy, University of Lethbridge, Lethbridge, AB T1K3M4, Canada
           \and
           D. Benielli \at
           Aix Marseille Universit\'e, CNRS, LAM (Laboratoire d'Astrophysique de Marseille) UMR 7326, 13388, Marseille, France
           \and
           T. Fulton \& P. Imhof \at
           Blue Sky Spectroscopy, Lethbridge, AB, T1J 0N9, Canada; \\
           Institute for Space Imaging Science, Department of Physics \& Astronomy, University of Lethbridge, Lethbridge, AB T1K3M4, Canada
           \and
           T. Konopczy\'nski \at
           Wroc{\l}aw University of Technology, 27 Wybrze\.ze Wyspia\'nskiego St, 50-370 Wroc{\l}aw, Poland\\
           Herschel Science Centre, ESAC, P.O. Box 78, 28691 Villanueva de la Ca\~nada, Madrid, Spain
           \and 
           T. Lim \at
           RAL Space, Rutherford Appleton Laboratory, Didcot OX11 0QX, UK
           \and
           N. Lu \at
           NHSC/IPAC, 100-22 Caltech, Pasadena, CA 91125, USA
           \and
           N. Marchili \at
           Universit\'a di Padova, I-35131 Padova, Italy
           \and
           D. Naylor \at
           Institute for Space Imaging Science, Department of Physics \& Astronomy, University of Lethbridge, Lethbridge, AB T1K3M4, Canada
           \and
           B. Swinyard \at
           RAL Space, Rutherford Appleton Laboratory, Didcot OX11 0QX, UK; \\
           Dept. of Physics \& Astronomy, University College London, Gower St, London, WC1E 6BT, UK
}

\date{Received: \today / Accepted: \today}

\maketitle

\begin{abstract}
We present a method to derive the relative pointing offsets for SPIRE Fourier-Transform Spectrometer (FTS) solar system object (SSO) calibration targets, which were observed regularly throughout the {\it Herschel} mission. We construct ratios $R_{\textrm{obs}}(\nu)$ of the spectra for all observations of a given source with respect to a reference. The reference observation is selected iteratively to be the one with the highest observed continuum. Assuming that any pointing offset leads to an overall shift of the continuum level, then these $R_{\textrm{obs}}(\nu)$ represent the relative flux loss due to mispointing. The mispointing effects are more pronounced for a smaller beam, so we consider only the FTS short wavelength array (SSW, 958--1546 GHz) to derive a pointing correction. We obtain the relative pointing offset by comparing $R_{\textrm{obs}}(\nu)$ to a grid of expected losses for a model source at different distances from the centre of the beam, under the assumption that the SSW FTS beam can be well approximated by a Gaussian. In order to avoid dependency on the point source flux conversion, which uses a particular observation of Uranus, we use extended source flux calibrated spectra to construct $R_{\textrm{obs}}(\nu)$ for the SSOs. In order to account for continuum variability, due to the changing distance from the \textit{Herschel} telescope, the SSO ratios are normalised by the expected model ratios for the corresponding observing epoch. We confirm the accuracy of the derived pointing offset by comparing the results with a number of control observations, where the actual pointing of \textit{Herschel} is known with good precision. Using the method we derived pointing offsets for repeated observations of Uranus (including observations centred on off-axis detectors), Neptune, Ceres and NGC\,7027. The results are used to validate and improve the point-source flux calibration of the FTS.
\end{abstract}

\section{Introduction}

The Spectral and Photometric Imaging REceiver (SPIRE: \citealt{griffin10}) is one of the three focal plane instruments on board the ESA \textit{Herschel} Space Observatory \citep{pilbratt10}. SPIRE consists of an imaging photometric camera and an imaging Fourier-Transform Spectrometer (FTS). The FTS works on the principle of interferometry: the incident radiation is separated in two beams that travel different optical paths before recombining. Thus the signal, that is measured by the FTS, is not the direct flux density in the passband, but the Fourier component of the spectral content (see \citealt{swinyard10,swinyard13} and the SPIRE Observers' Manual\footnote{The SPIRE Observers' Manual is available at the \textit{Herschel} Science Centre web:\\ \url{http://herschel.esac.esa.int/Docs/SPIRE/pdf/spire_om.pdf}} for more details). The final FTS spectra, after the inverse Fourier transform, cover two wide frequency bands:  SLW (447--990 GHz) and SSW (958--1546 GHz) at high (HR, $\Delta\nu=1.2$ GHz) and low (LR, $\Delta\nu=25$ GHz) spectral resolution.

The SPIRE FTS flux calibration is a two stage process (see \citealt{swinyard10,swinyard13} for more details): firstly, the telescope is used to derive the extended source calibration, which converts the interferogram signal timeline from Volts to W m$^{-2}$ Hz$^{-1}$ sr$^{-1}$ and provides level-1 spectra. Secondly, extended data is converted to flux density units of Jy ($10^{-26}$ W m$^{-2}$ Hz$^{-1}$) to provide point source calibrated (level-2) spectra. FTS uses Uranus as the primary point source calibrator.

The FTS flux calibration is split into two epochs, due to a change in the position of the internal beam-steering mirror (BSM) at the start of the {\it Herschel} operational day (OD) \texttt{1011} (19 Feb 2012, see \citealt{swinyard13}). The change effectively moved the centre of the FTS beam closer to the telescope optical axis. The earlier BSM position was at $1.7\arcsec$ with respect to the telescope commanded sky position. Therefore the currently implemented point-source flux calibration uses two sets of Uranus observations: ``before'' and ``after'' \od{1011}.

It is important to know the pointing offset for the primary calibrator. There is no \textit{a priori} knowledge of the actual pointing of the \textit{Herschel} telescope for staring observations with the FTS, i.e. in sparse mode. We know the commanded pointing error (or the absolute pointing error, APE) of \textit{Herschel} is $\sim 2-3\arcsec$ at 68\% confidence level throughout the \textit{Herschel} mission, with significant improvement for observations after 11 Mar 2013 (or \textit{Herschel}'s operational day \od{1032}), down to APE $\approx 0.8\arcsec$ (see \citealt{miguel} for details). The uncertainty in the pointing can lead to significant differences in the derived flux calibration, if a perfectly centred source is assumed. An additional source of pointing uncertainty is the stability of the pointing (or the relative pointing error RPE) once the telescope is commanded to the target position. The RPE is estimated at $0.3\arcsec$ for moving targets  (i.e. solar system objects, SSO) and smaller for non-moving ones \citep{miguel}. This is insignificant compared to the smallest SPIRE FTS beam ($17\arcsec$ at 1500 GHz, see later) and for the rest of the analysis we consider the pointing to be perfectly stable. Nevertheless, we include the RPE in the error budget for the results.

We have determined the pointing offsets for all Uranus observations, including those currently used for the point-source calibration.  Hence the presented work has important consequences for the flux calibration scheme as well as for future improvements by combining more than one observation of the primary calibrator.

Adding more calibration targets with repeated observations is a consistency check for the method itself and for the derived point source flux calibration scheme, especially for targets with relatively well known and accurate models. For example, without knowing the Neptune pointing we cannot conclude if the flux calibration scheme is good, because any pointing offset of Neptune will lead to a difference of the observed versus the model flux of Neptune, which will not be related to the accuracy of the flux calibration scheme.

The structure of this paper is the following: in the next section (\S~\ref{sec:method}) we present the method, using two types of synthetic source models -- a Dirac $\delta$-function or a disk with a given radius and we derive a grid of the expected flux losses for a source at a range of distances from the centre of the FTS Gaussian beam. Then in \S~\ref{sec:results} we present the results on the relative pointing offsets for Uranus, Neptune, Ceres and NGC\,7027, concentrating mostly on Uranus as the primary calibrator for the SPIRE Spectrometer. We end the paper in \S~\ref{sec:conclusions} with our conclusions and some insights on possible further improvements in the flux calibration scheme.

\section{Method}
\label{sec:method}

The method uses level-1 data of repeated FTS observations of calibration targets, performed in high (HR) or low (LR) spectral resolution. All level-1 data were reprocessed with the latest user reprocessing script in the Herschel Interactive Processing Environment (\textsc{hipe}, \citealt{hipe}) version 11 and SPIRE calibration tree version \textsc{spire\_cal\_11\_0}. To avoid any dependence on Uranus, we use level-1 data, because level-2 spectra are calibrated using a particular Uranus observation. In addition, we consider only an overall continuum shift and ignore any change in the spectral slope or other non-linear effects. 

\begin{figure}
\centerline{
\includegraphics[width=0.7\textwidth]{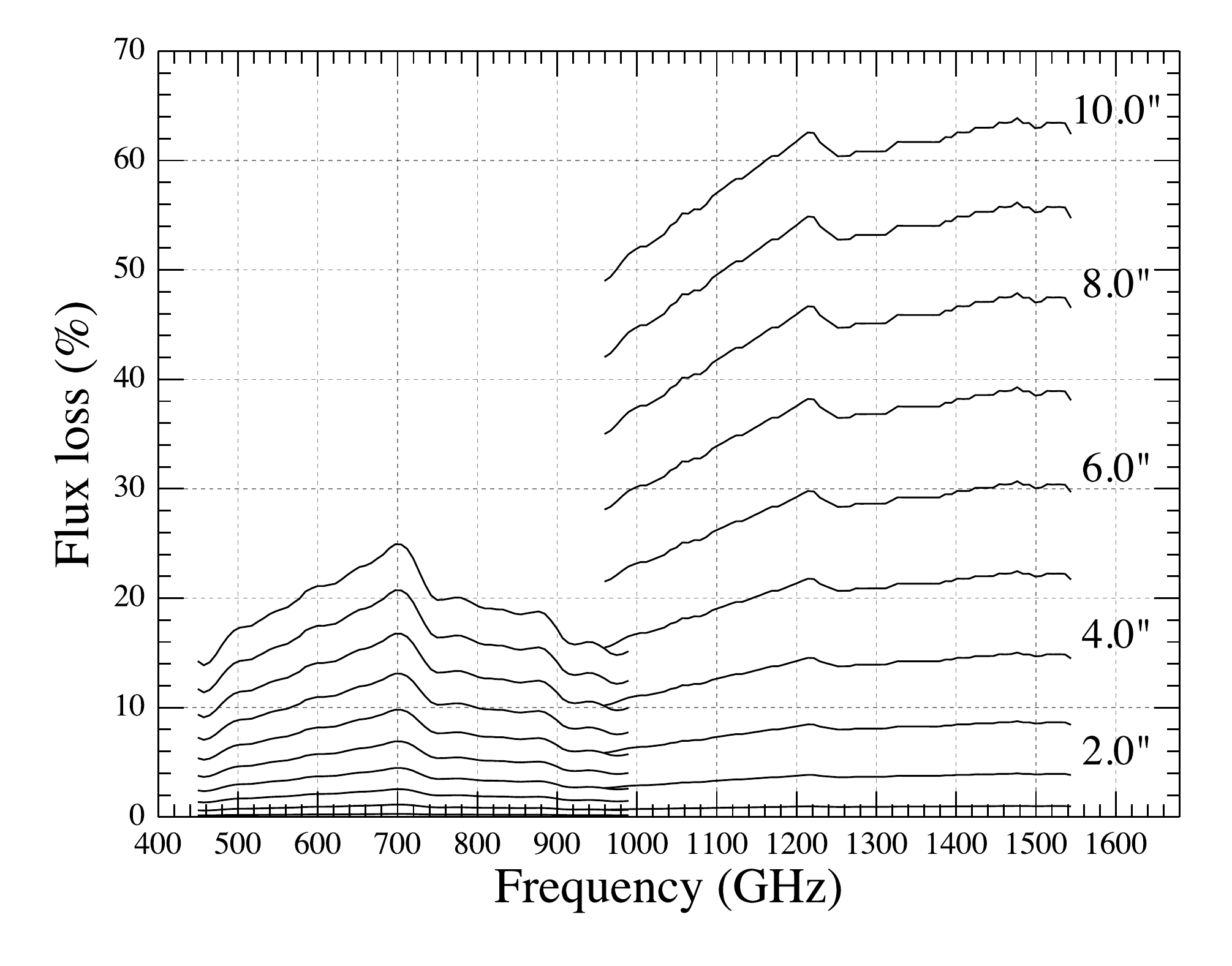}}
\caption{The flux loss in \% for a perfect point source (Dirac $\delta$-function) placed at different offsets from the FTS beam centre. These curves were derived assuming a Gaussian SPIRE FTS beam response at each frequency. This is a good approximation for SSW (900--1500 GHz) \citep{gibion13}.}
\label{loss}
\end{figure}

The analysis is performed solely using the FTS short wavelength array central detector, SSWD4. A simple Gaussian for the beam provides a good approximation over the frequencies of interest, 900--1500 GHz (see \citealt{gibion13} for more details) where the full width at half maximum (FWHM) is of the order of 17$\arcsec$. A smaller beam, however, means any pointing offset leads to more significant flux loss. This is shown in Fig.~\ref{loss} for a perfect point source (a $\delta$-function) placed at different offsets from the FTS beam centre. And indeed, the flux loss for SLW even at the extreme $10\arcsec$ offset is 15-25\%, while it is 50-60\% in SSW. 

As seen by \textit{Herschel}, Uranus, Neptune and Ceres are variable sources due to their distance from the Sun, their rotation and the telescope position on its orbit around the second Lagrange point L2. The changes in the brightness due to the variability will have the same effect as a pointing offset, i.e. overall continuum shift. This complicates the analysis because we need to know the predicted planet flux for the exact time of the observation. The planetary models used in this analysis are ESA-4 for Uranus (based on the updated model of \citealt{uranus}) and ESA-3 for Neptune (based on the updated model of \citealt{neptune})\footnote{Both planets models are available at: \url{ftp://ftp.sciops.esa.int/pub/hsc-calibration/PlanetaryModels/}}. The absolute systematic flux calibration uncertainties  in these models, derived by comparing Uranus and Neptune, are of the order of 3\% \citep{swinyard13}. For Ceres we use the models by \citet{muller}, which are more uncertain, at $\sim10$\%, because of the systematic phase variance due to the use of shorter wavelength data from ISO.


To derive the flux loss we assume the FTS has a Gaussian beam $G_\mathrm{B}(x,\theta_\mathrm{B})$ with FWHM $\theta_\mathrm{B}(\nu)$ at a given frequency $\nu$. We calculate the flux loss for two cases of a source shape: (1) a perfect point source represented by a Dirac $\delta$-function and (2) a disk with radius $\theta_\mathrm{p}$, which in the 1D case is a tophat function: 

\begin{equation}
h(x,\theta_{\mathrm{p}}) = 
\begin{cases} 
  1,  & |x| \leq \theta_{\mathrm{p}} \\
  0,  & \mathrm{else} 
\end{cases}
\end{equation}

\begin{figure*}
\centerline{
\includegraphics[width=0.49\textwidth]{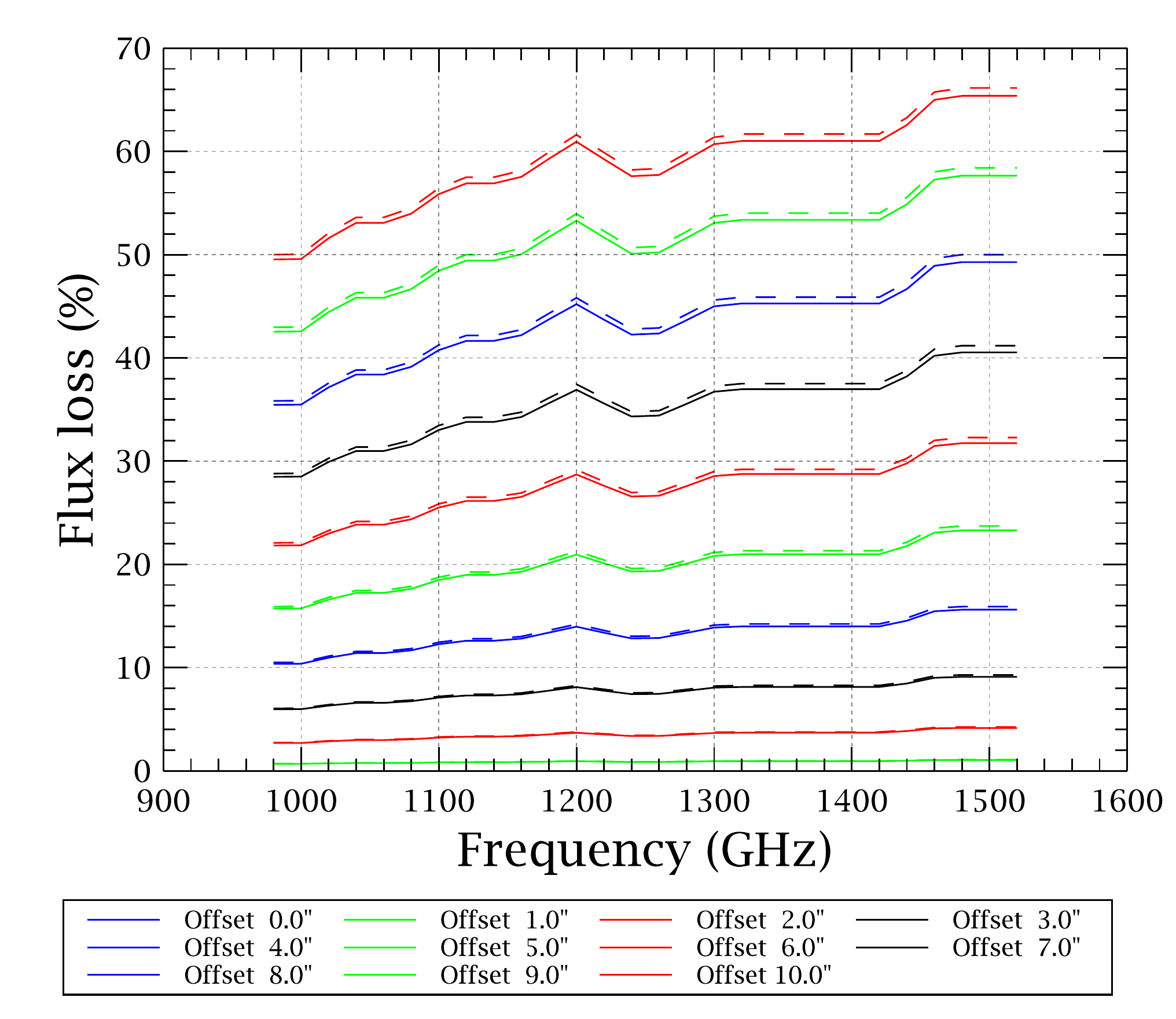}
\includegraphics[width=0.49\textwidth]{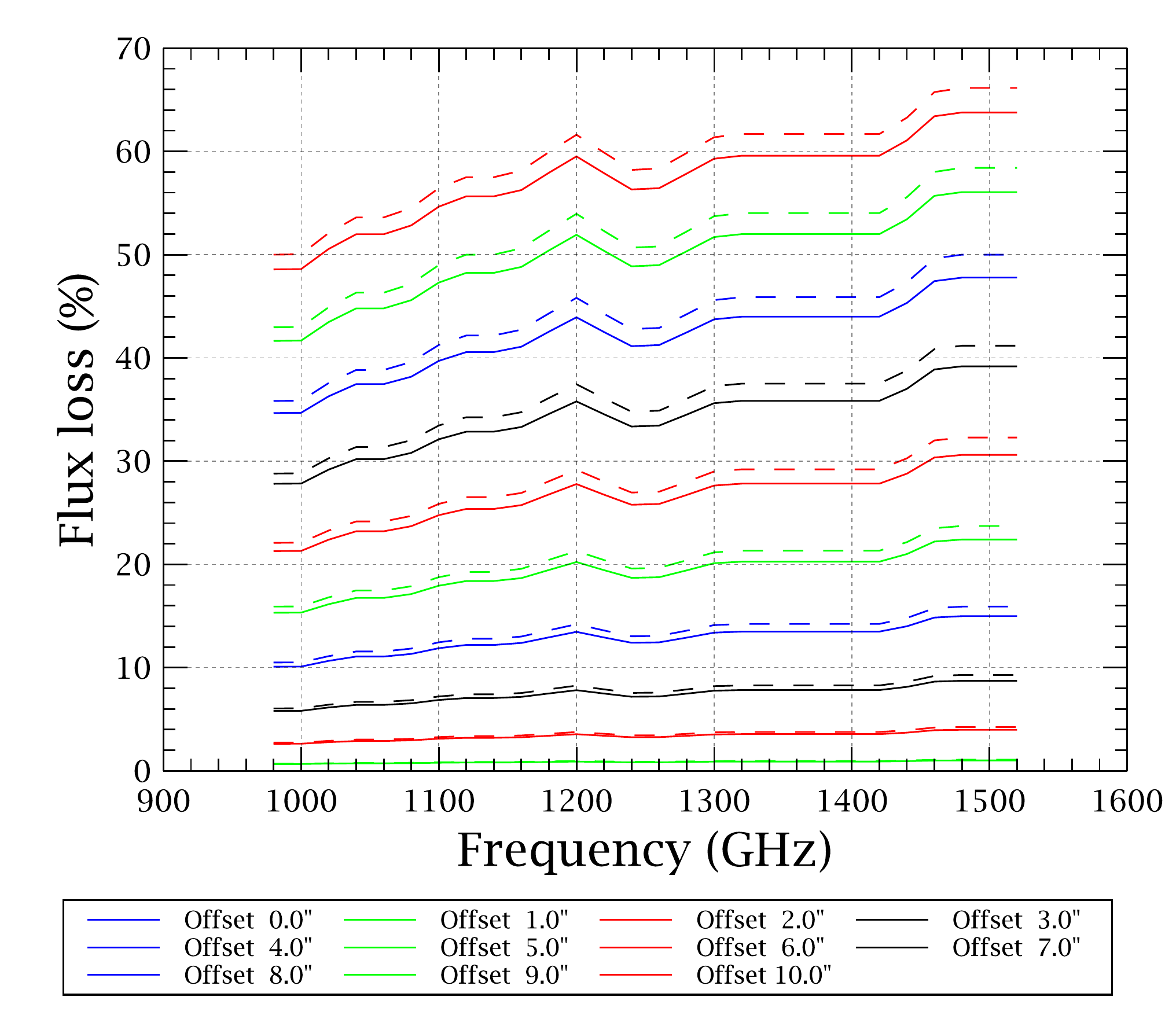}}
\caption{Flux loss in SSW assuming $\delta$-function (dashed lines) and a tophat disk (continuous lines) with a fixed radius of $1.7\arcsec$ (typical Uranus angular radius) on the left and a more extreme case of a synthetic source having a tophat disk of $3\arcsec$ radius on the right.}
\label{fig_losses}
\end{figure*}

In both cases, the source is moved on a grid of pointing offsets $\alpha$ from 0 to a maximum $10\arcsec$ with respect to centre of the Gaussian beam and then it is convolved with the beam. That is

\begin{equation}
c_\alpha(\nu) = \int\limits_{-\infty}^{+\infty} G_\mathrm{B}(t,\theta_{\mathrm{B}}) \delta(\alpha - t) dt = G_\mathrm{B}(\alpha,\theta_{\mathrm{B}}),
\label{eq_delta}
\end{equation}
for $\delta(\alpha) = 1$ source, and 

\begin{equation}
c_{\alpha}(\nu) = \int\limits_{-\infty}^{+\infty} G_\mathrm{B}(t,\theta_{\mathrm{B}}) h(\alpha - t) dt = 
 C\,\left[\mathrm{erf}(r+) - \mathrm{erf}(r-)\right],
 \label{eq_tophat}
\end{equation}
for a uniform disk of radius $\theta_\mathrm{p}$, where $r\pm = \sqrt{4\,\ln{2}}\,(\alpha \pm \theta_{\mathrm{p}})/\theta_\mathrm{B}$,  $C$ is a normalisation constant and $\mathrm{erf}(x) = \frac{2}{\sqrt{\pi}}\,\int_0^x e^{-t^2} dt$ is the error function. For a uniform disk centred in the beam, i.e. $\alpha=0$, then $c_0$ is the beam correction factor $K_\mathrm{beam}$ as used in the flux calibration scheme \citep{swinyard13}. 
The flux loss at a given frequency is then
\begin{equation}
f_{\alpha}(\nu) = 1 - c_\alpha(\nu)/c_0(\nu),
\end{equation}
where $c_\alpha$ comes either from Eq.~\ref{eq_delta} or Eq.~\ref{eq_tophat}. The results are shown in Fig.~\ref{fig_losses}. In the case of Uranus, with a typical size of $\theta_\mathrm{p} = 1.7\arcsec$, the tophat and the $\delta$-function results are different by less than 1\%, even for an extreme pointing offset of $10\arcsec$. For Neptune, which has an angular radius smaller than Uranus -- typically around $1.1\arcsec$ -- this difference is negligible.

Let us assume a reference observation \textsc{<ref>} with the source exactly in the centre of the beam. If the source has constant brightness then 
\begin{equation}
R_{obs}(\nu) = 1 - \frac{\textrm{level-1}(\textsc{obs})}{\textrm{level-1}(\textsc{ref})}
\end{equation}
will be a measure of the flux loss at frequency $\nu$, and the loss can be considered as only due to the pointing offset of \textsc{<obs>} with respect to \textsc{<ref>}. A source with constant brightness included in this study is NGC~7027 but for the solar system objects Uranus, Neptune and Ceres, the brightness depends on when they were observed, which means the direct ratio $R_{obs}(\nu)$ has to be corrected for the predicted flux from the planet, i.e.
\begin{equation}
R_{obs}(\nu) = 1 - \frac{\textrm{level-1}(\textsc{obs})}{\textrm{level-1}(\textsc{ref})} \times R_{\textrm{model}},
\label{eq_robs_model}
\end{equation}
where $R_{\textrm{model}} = \textrm{model}(\textsc{ref})/\textrm{model}(\textsc{obs})$ is the ratio of the modelled fluxes at the two epochs of \textsc{<ref>} and \textsc{<obs>}. Note that the model ratio $R_{\textrm{model}}$ is independent of the frequency, i.e. the models for observations at different epochs only differ by an overall continuum shift, independent of the frequency.

The choice of the reference observation is iterative. All $R_{obs}(\nu)$ ratios are constructed taking an arbitrary \textsc{<ref>} observation and then we pick the one that has the lowest $R_{obs}(\nu)$, below or at zero. Indeed, if the selected reference observation is the one where the target is closest to the centre of the beam then there will be no $R_{obs}$ smaller than zero, within the margin of the model uncertainties.

An HR observation is always selected as reference. In order to avoid large fluctuations in the ratios $R_{obs}(\nu)$ we smooth it with a 20\,GHz boxcar function, after performing the division. For LR observations we interpolate the smoothed reference HR spectra to the same frequency grid as the LR observation, before deriving the ratio. This avoids huge fluctuations, especially in the case of Neptune, where there are emission and absorption lines present in the spectra.

\begin{figure*}
\centerline{
\includegraphics[width=0.5\textwidth]{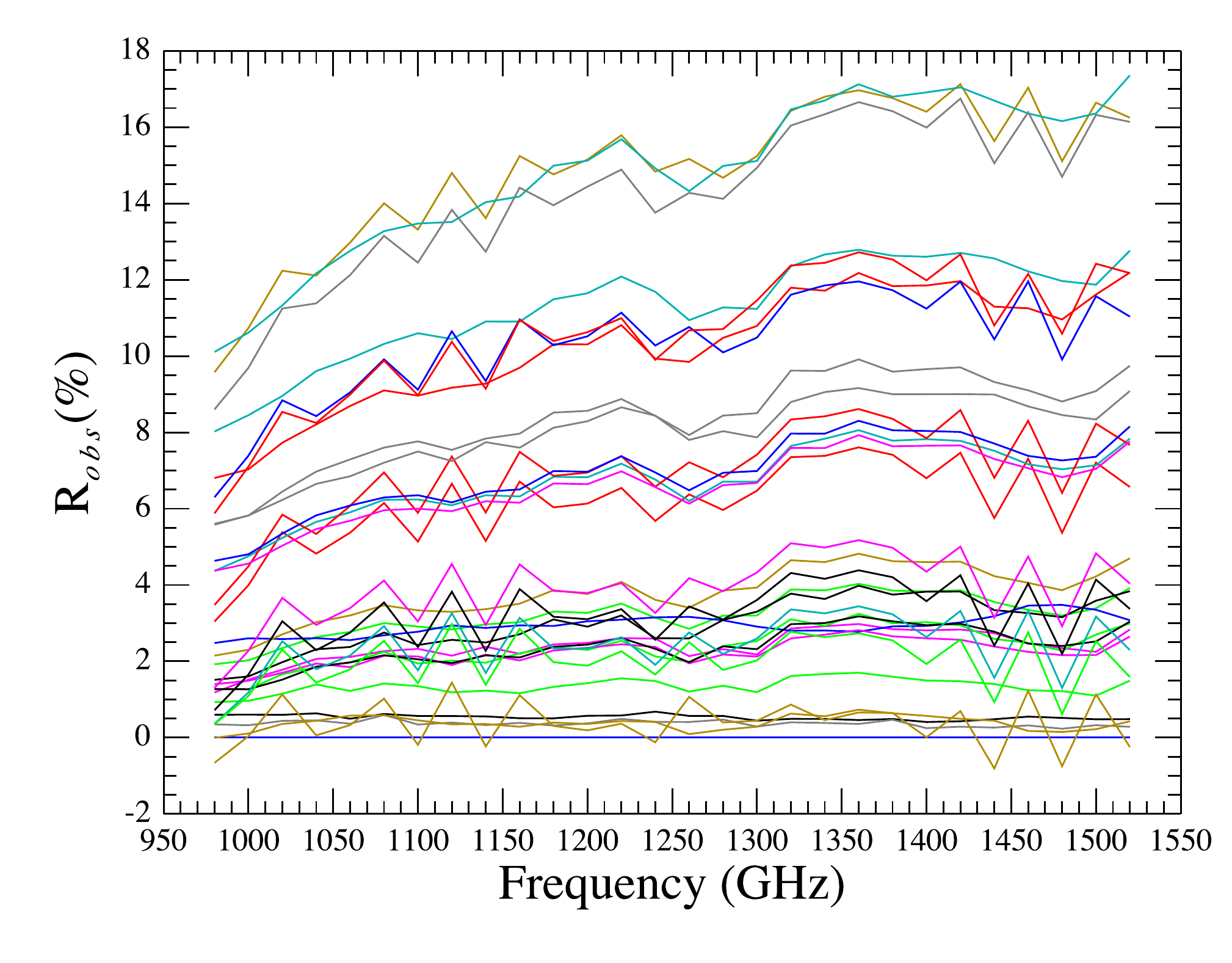} \hfill
\includegraphics[width=0.5\textwidth]{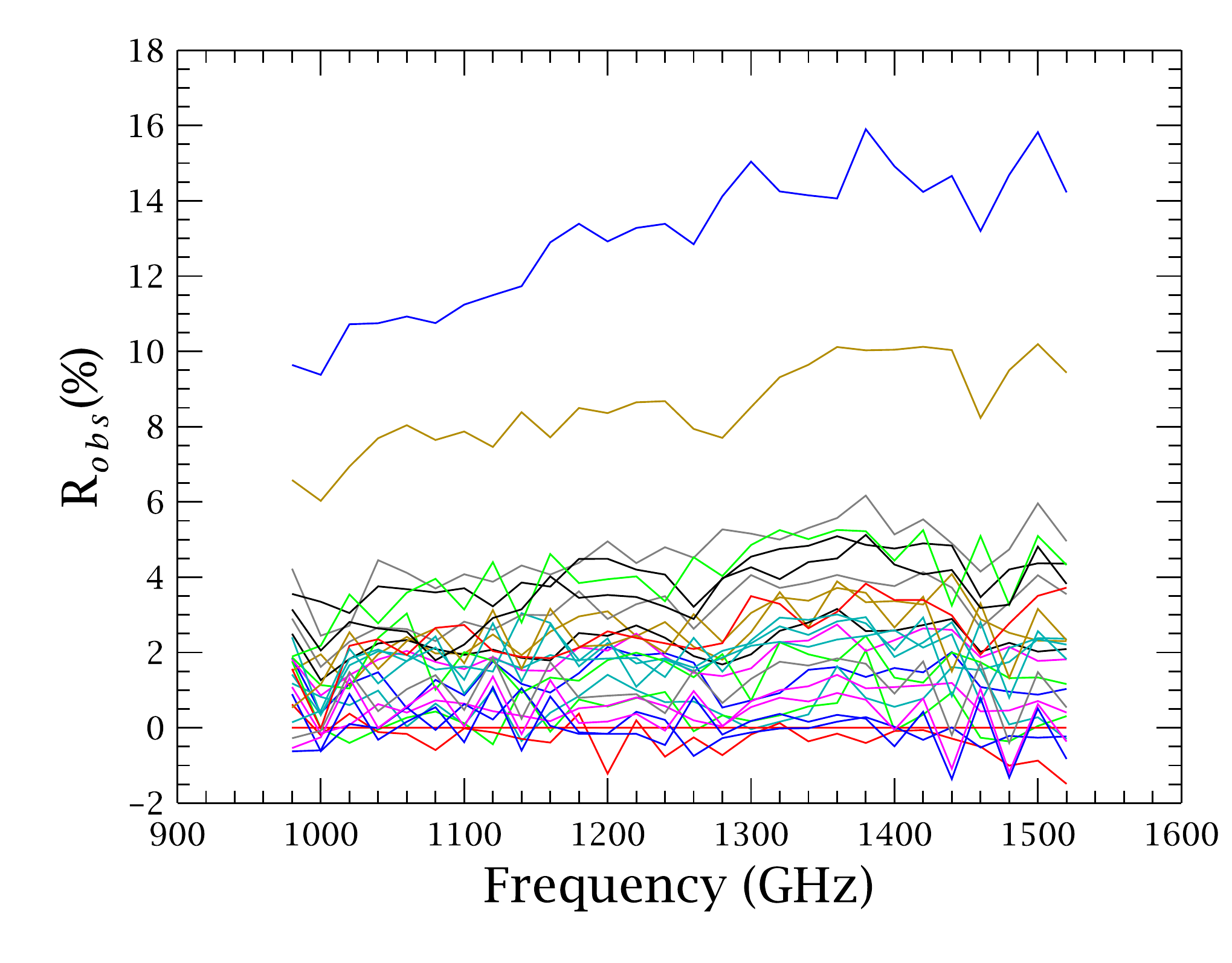}}
\centerline{
\includegraphics[width=0.5\textwidth]{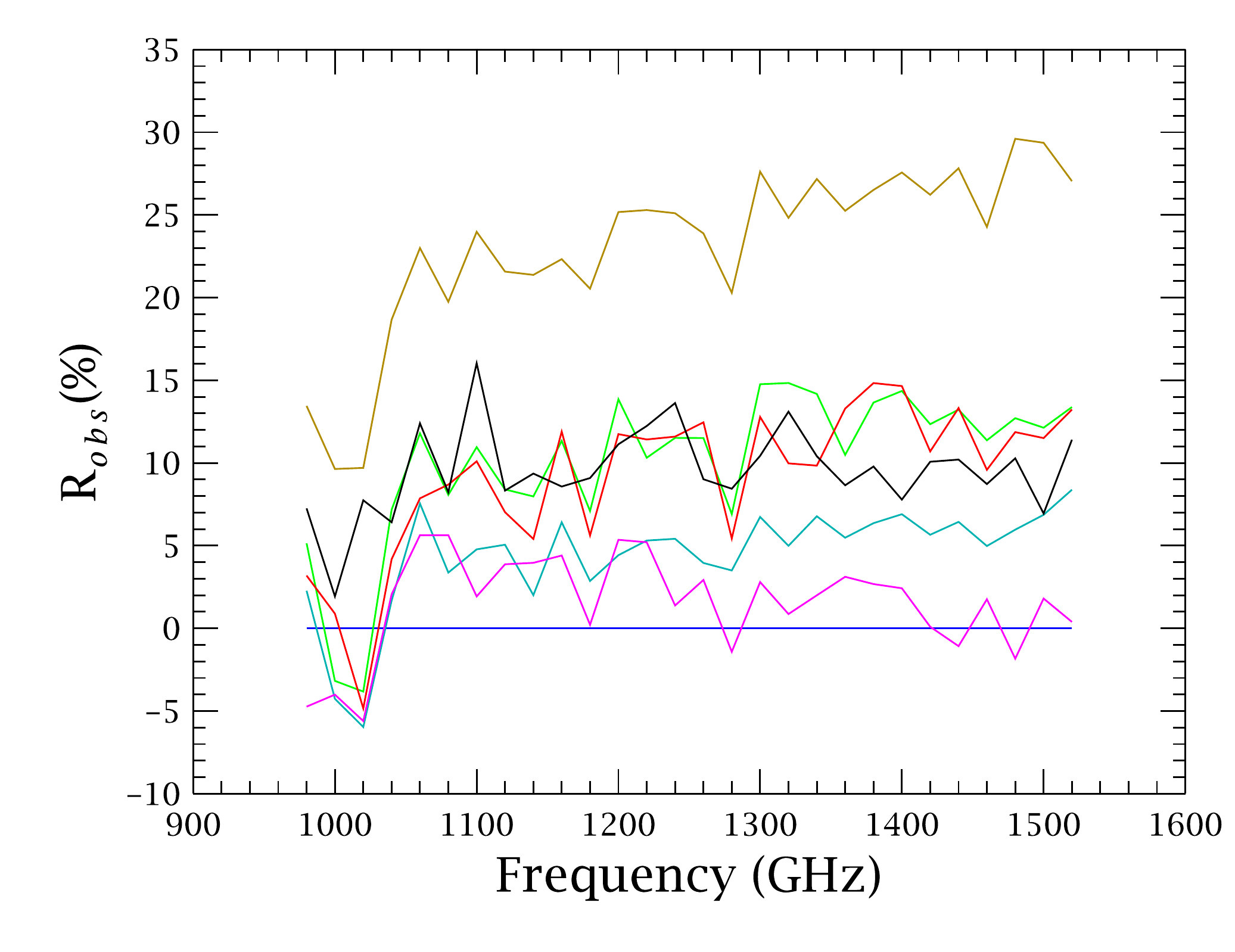} \hfill
\includegraphics[width=0.5\textwidth]{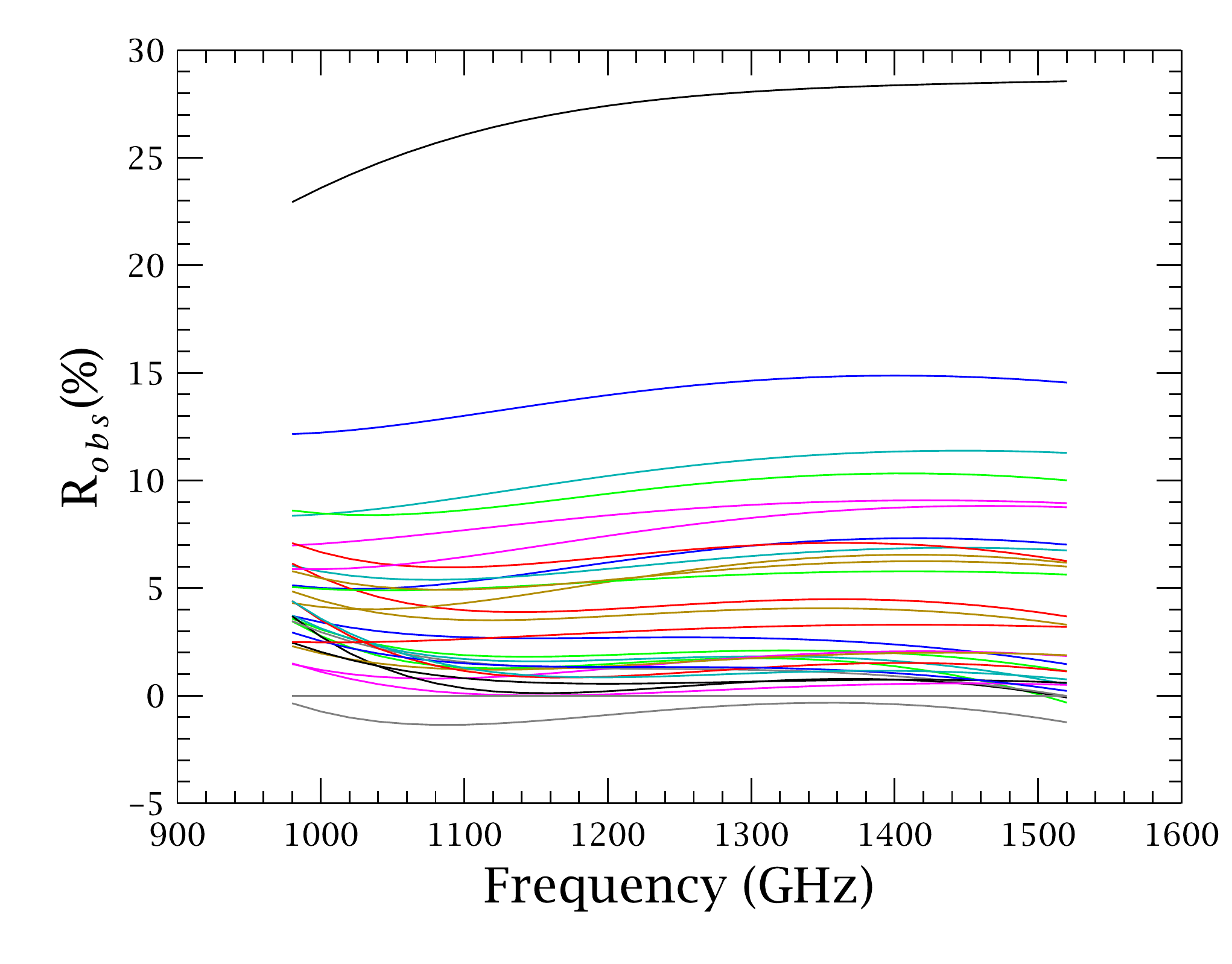}}
\caption{The derived $R_{obs}$ for the calibration targets included in this study. Uranus (upper left), Neptune (upper right), Ceres (lower left) and NGC~7027 (lower right). The different colours are used for the different observations. }
\label{fig_robs}
\end{figure*}

\begin{figure*}
\centerline{
\includegraphics[width=0.5\textwidth]{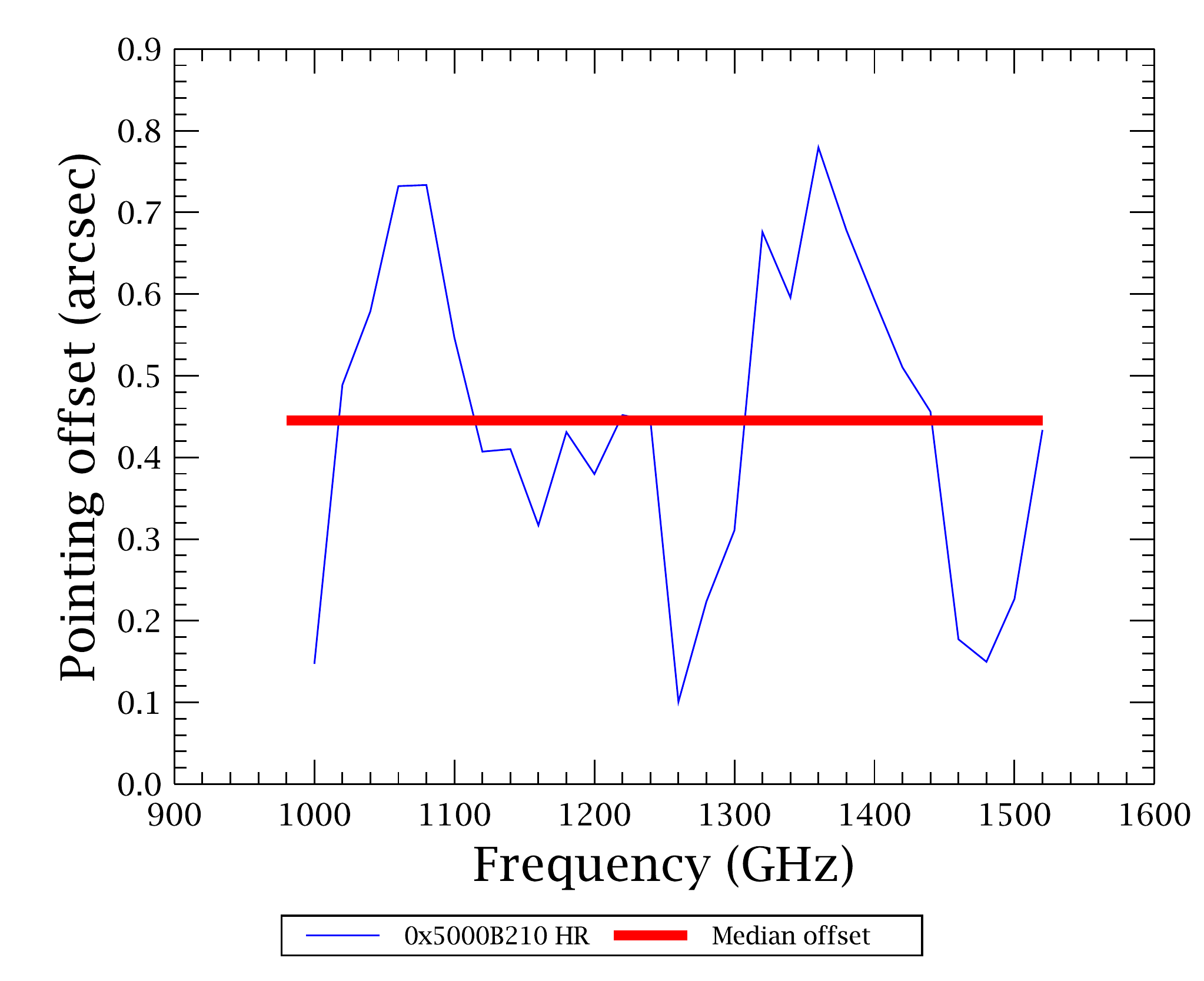} \hfill
\includegraphics[width=0.5\textwidth]{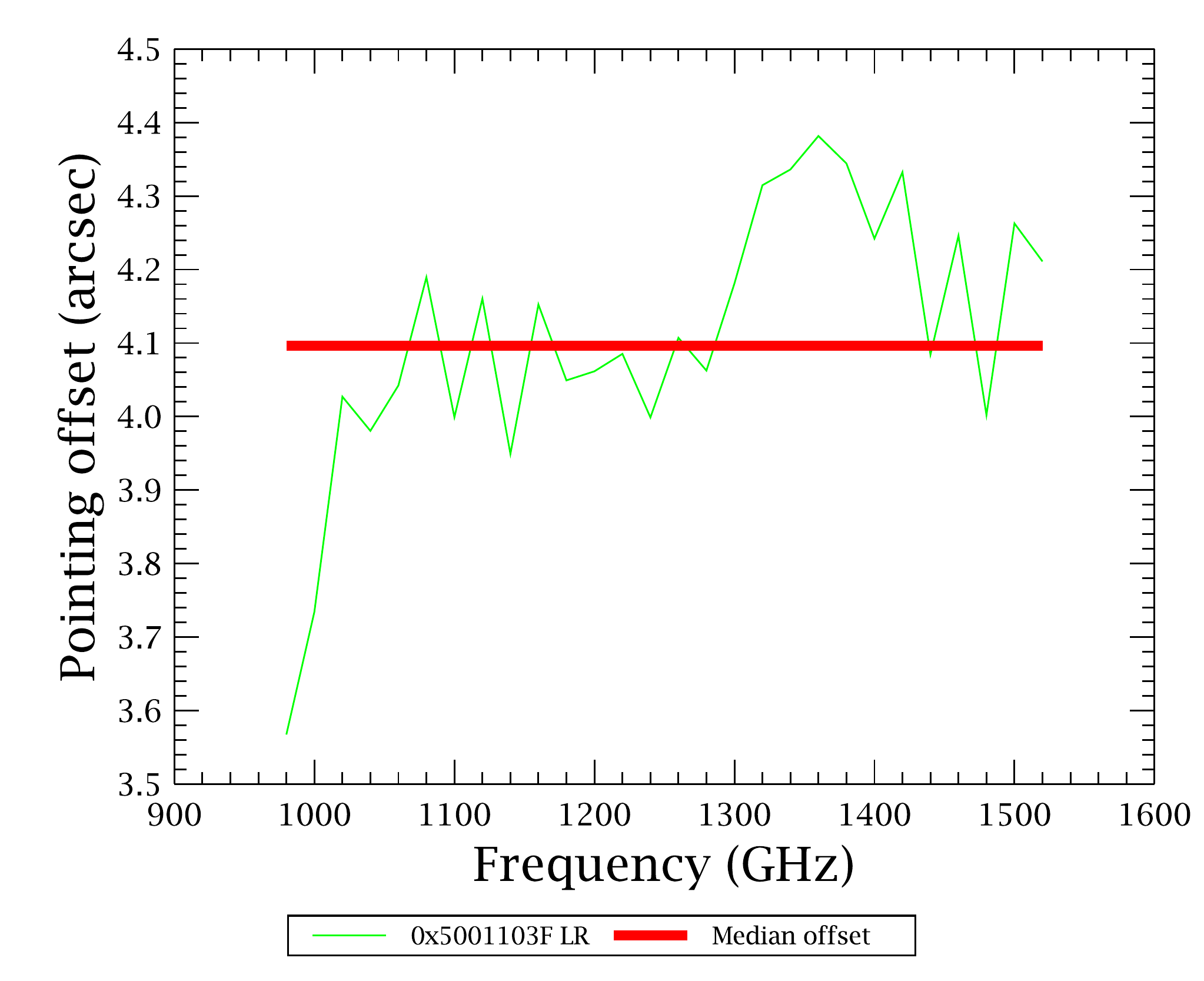}}
\centerline{
\includegraphics[width=0.5\textwidth]{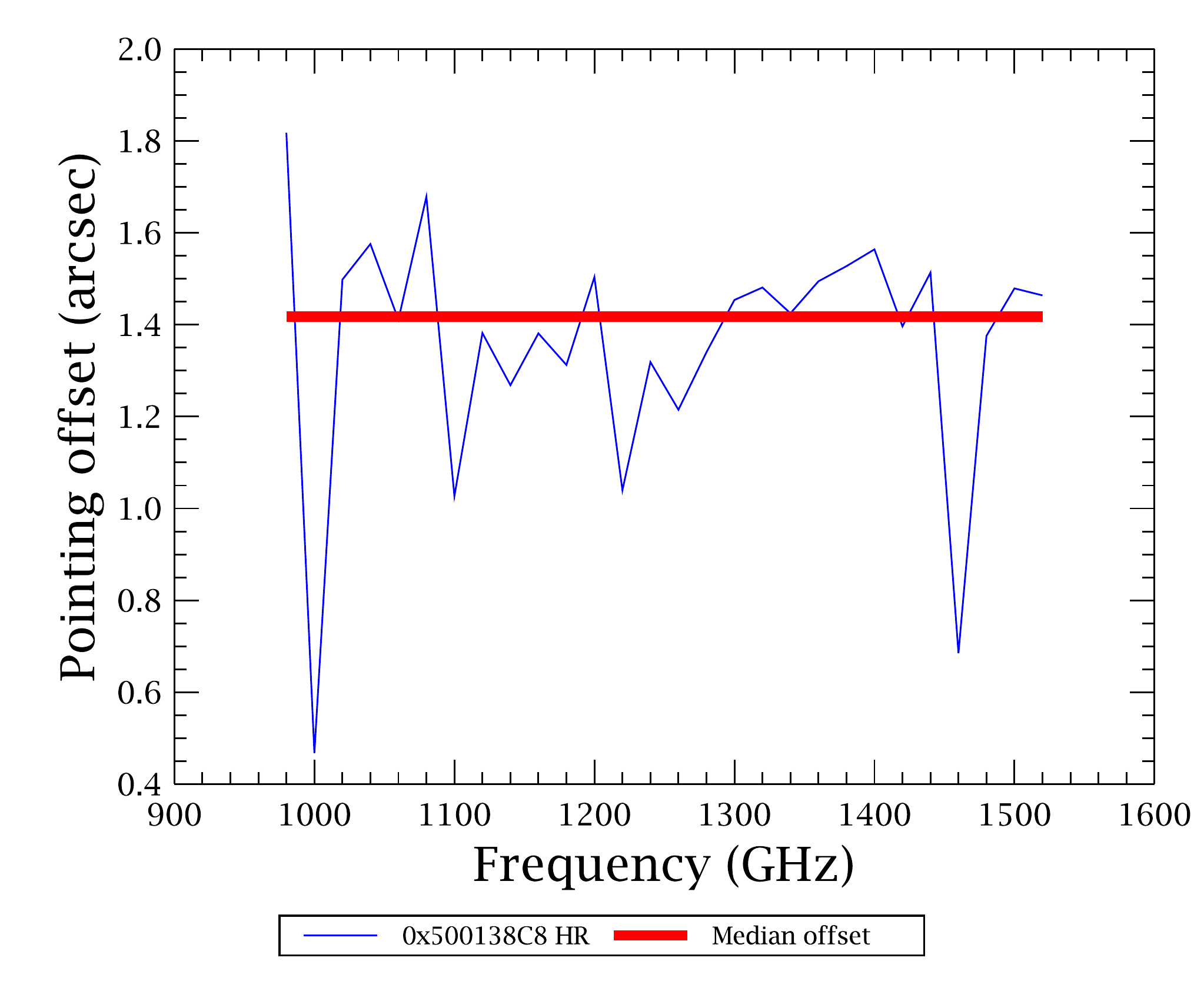} \hfill
\includegraphics[width=0.5\textwidth]{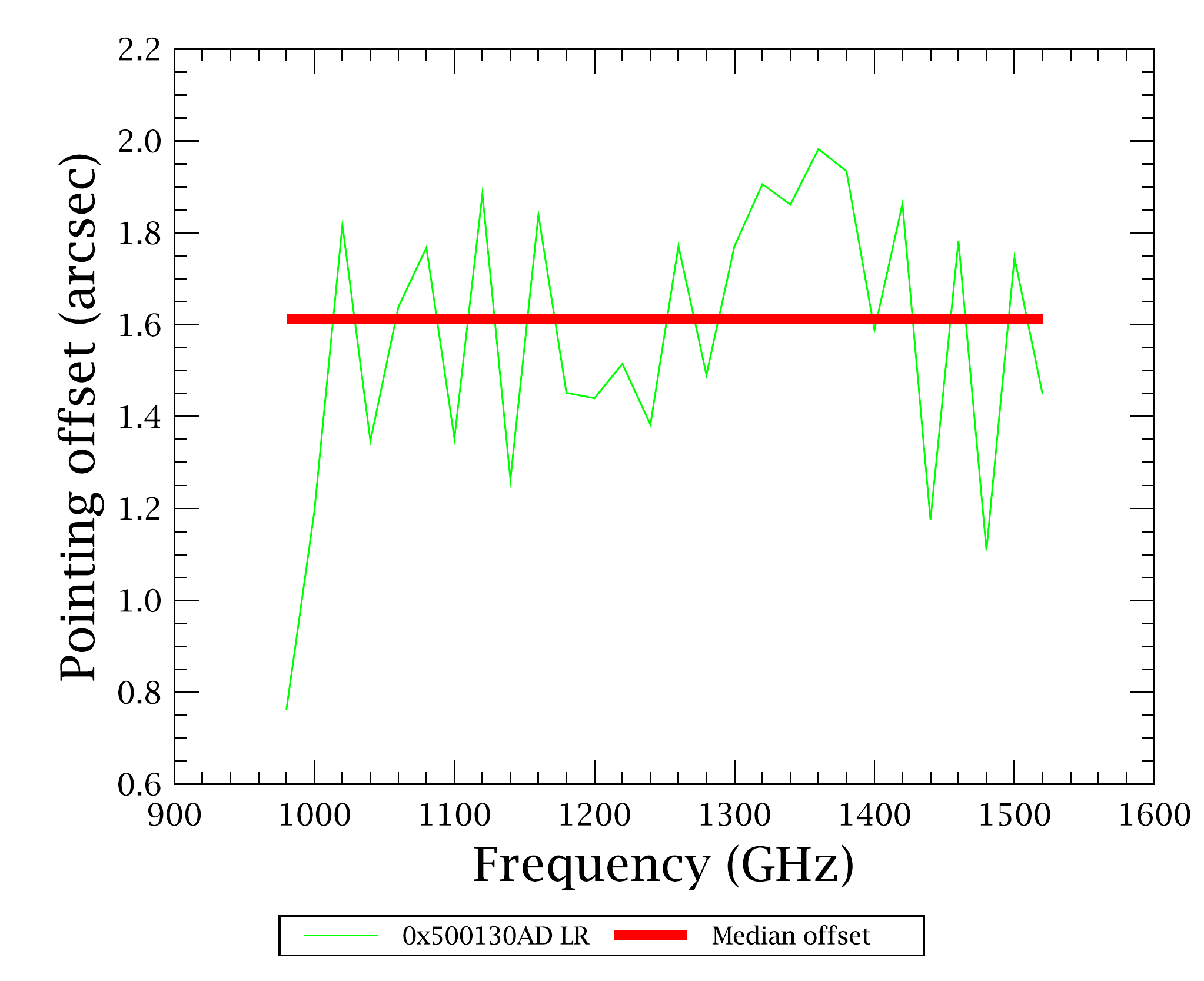}}
\centerline{
\includegraphics[width=0.5\textwidth]{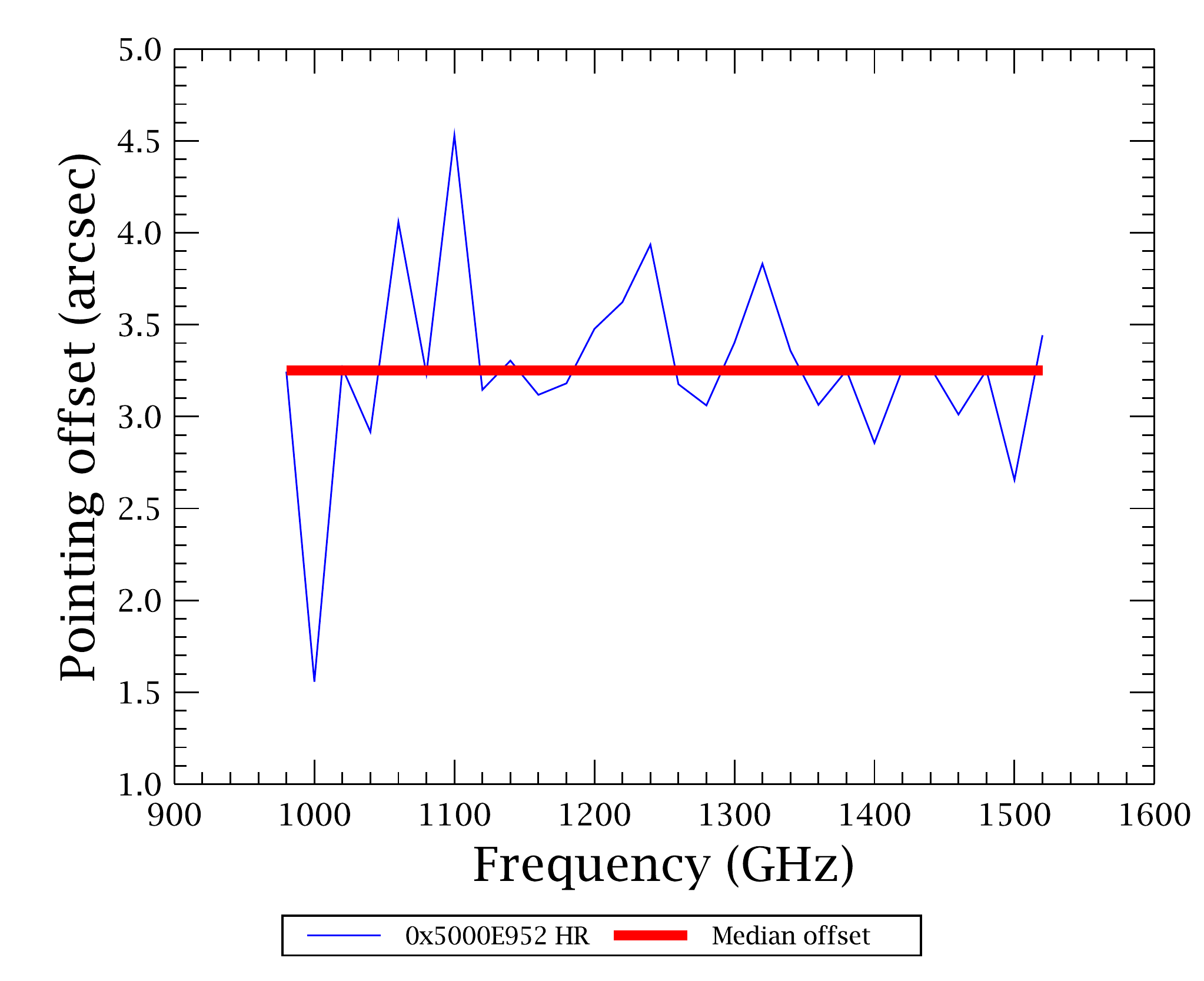} \hfill
\includegraphics[width=0.5\textwidth]{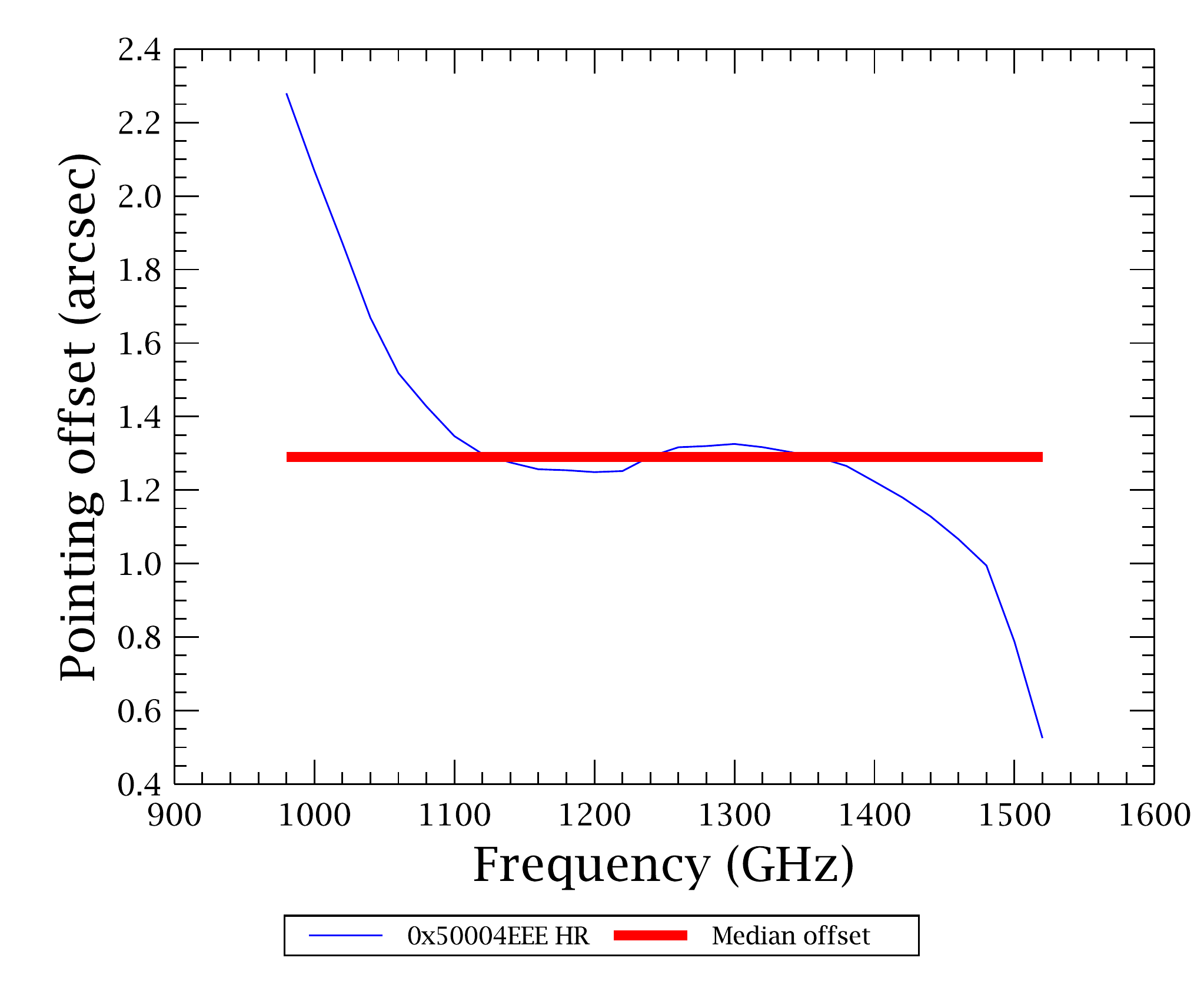}}
\caption{Examples for the derived pointing offsets for Uranus HR and LR (top), Neptune HR and LR (middle), Ceres HR (bottom left) and NGC~7027 HR (bottom right). In all cases the thick red line is the derived median pointing offset, calculated using frequencies from 1100 to 1400 GHz.}
\label{fig_examples}
\end{figure*}

Fig.~\ref{fig_robs} shows the $R_{obs}(\nu)$ for all targets included in this study. The \textit{relative pointing offset} $\alpha$\ is derived by comparing $R_{obs}(\nu)$ with the grid of simulated offsets $f_{\alpha}(\nu)$ (i.e. Fig.~\ref{fig_losses}). If the flux loss is only due to pointing then the derived pointing offset $\alpha$ should be constant with frequency. The results for a few observations are shown in Fig.~\ref{fig_examples}. We calculate the median offset and the median absolute deviation (as a robust measure of the spread) considering frequencies from 1100 to 1400 GHz,  in order to avoid large fluctuations near the band edges.

As the method is relative, the uncertainties coming from the models (for SSOs) cancel out and the only factors entering the error budget are the pointing stability (the RPE) of $0.3\arcsec$ (assumed perfect for the method) and the uncertainty of 1\% on $R_{obs}(\nu)$ coming from the statistical repeatability of the observations \citep{swinyard13}. In most cases the RPE is the dominant uncertainty in the derived offsets.

\section{Results}
\label{sec:results}

\begin{figure*}
\centerline{
\includegraphics[width=0.5\textwidth]{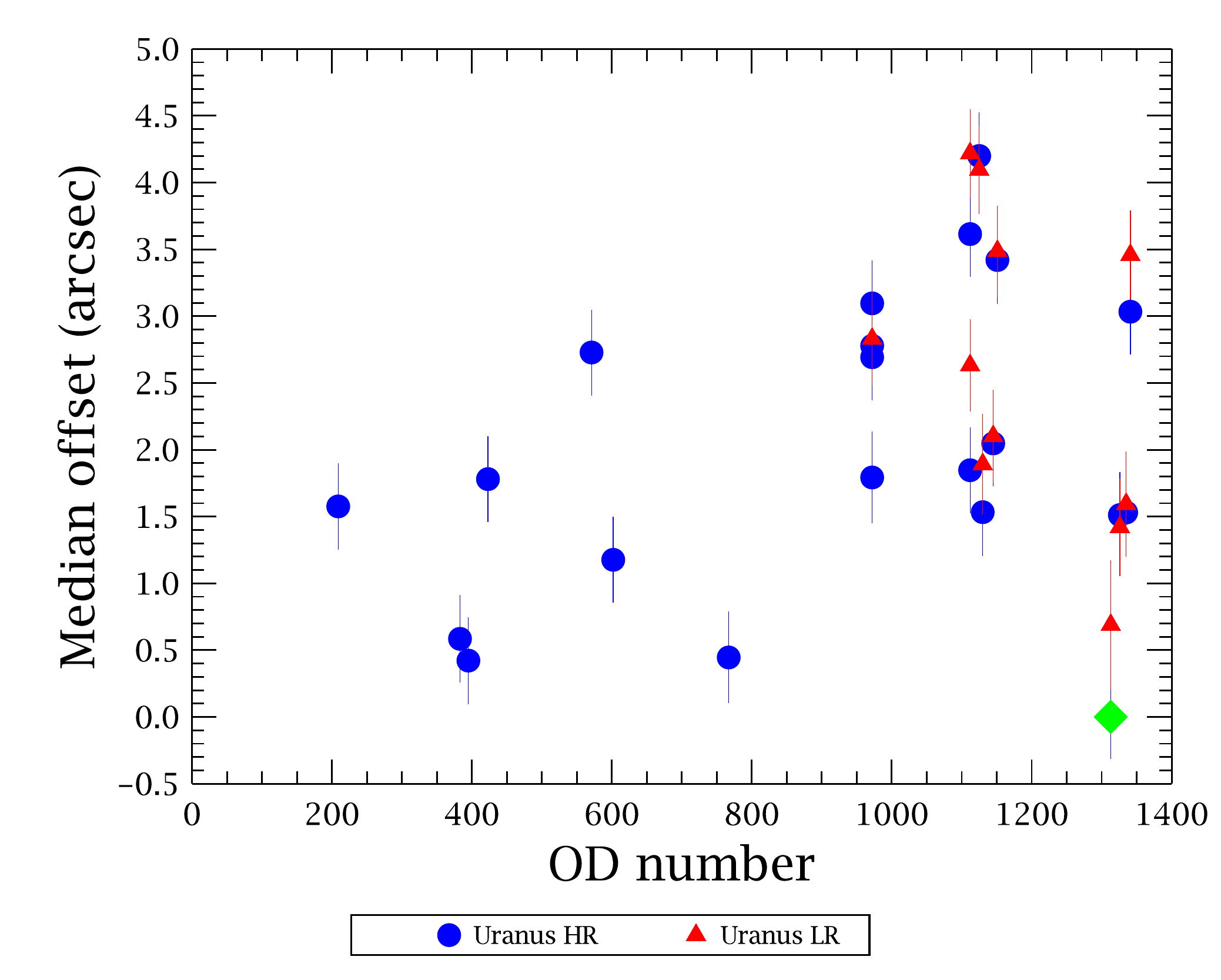} \hfill
\includegraphics[width=0.5\textwidth]{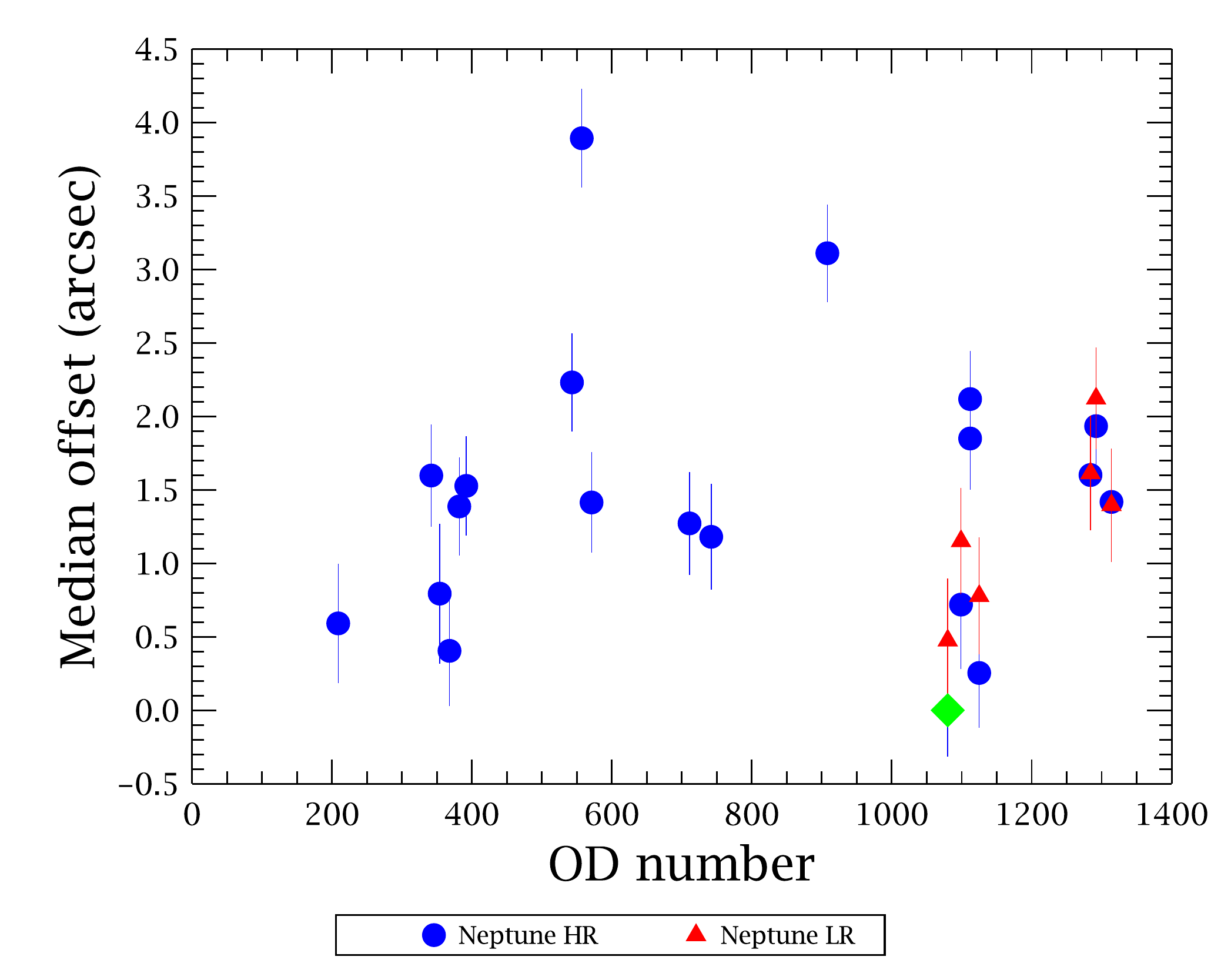}}
\centerline{
\includegraphics[width=0.5\textwidth]{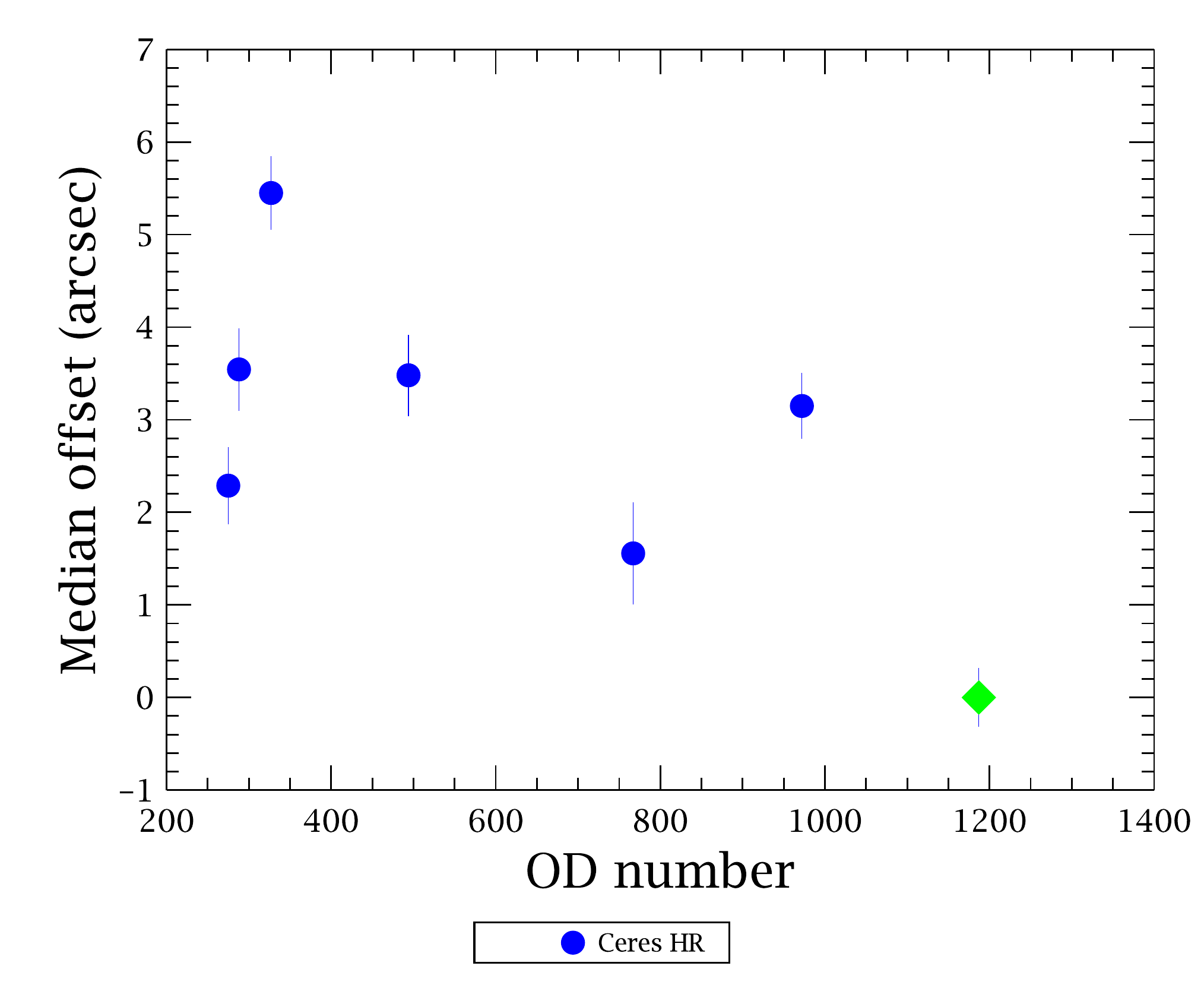} \hfill
\includegraphics[width=0.5\textwidth]{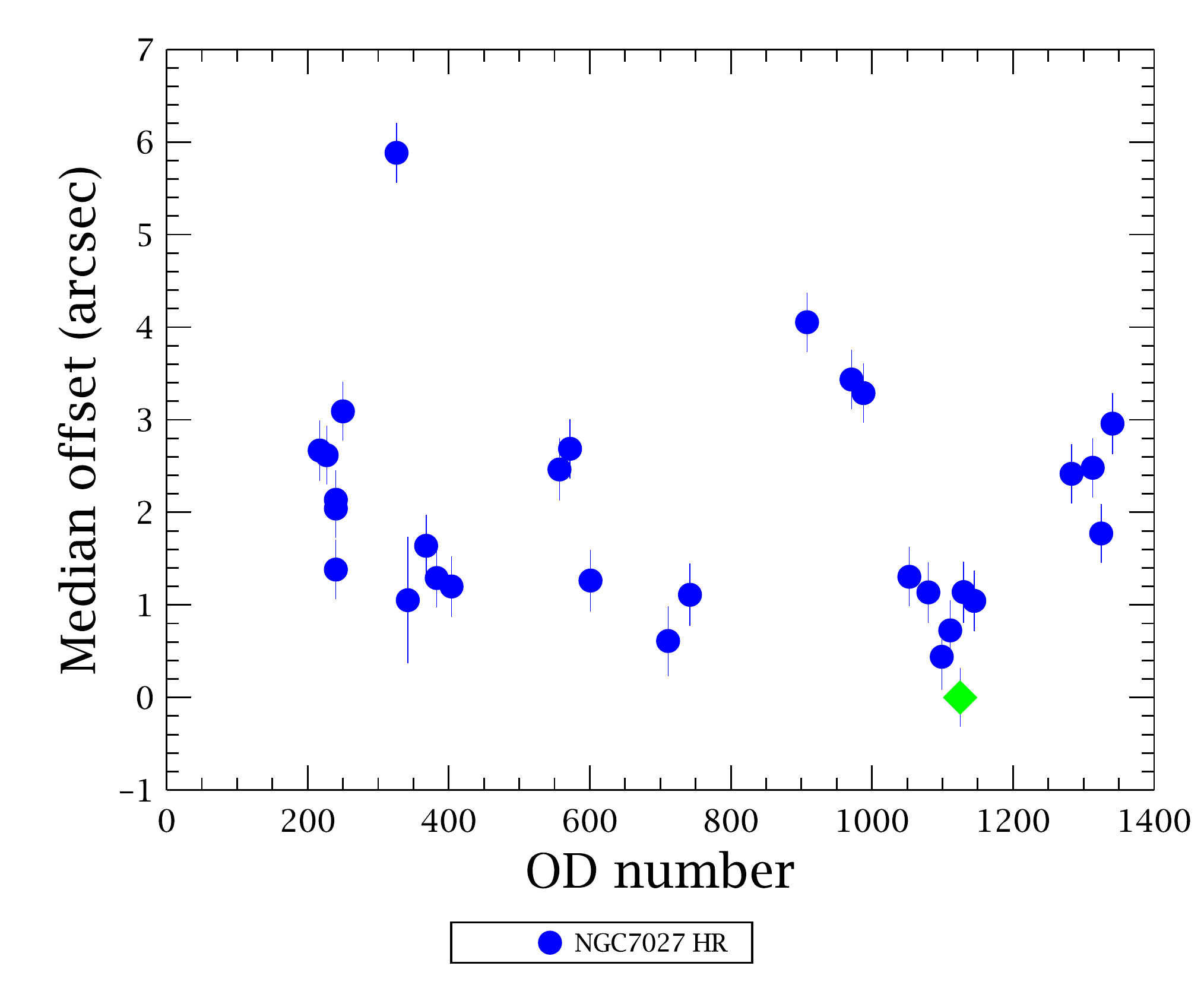}}
\caption{The derived median pointing offsets for Uranus (top left), Neptune (top right), Ceres (lower left) and NGC~7027 (lower right). The error bars are the sum (in quadrature) of the median absolute deviation of the pointing offset and the pointing jitter of $0.3\arcsec$. The reference observations are indicated with a green diamond. The blue points are for HR observations, the red triangles are for LR (only Uranus and Neptune).}
\label{fig_results}
\end{figure*}

The above method was applied to all observations of Uranus, Neptune, Ceres and NGC~7027 and the results are shown in Fig.~\ref{fig_results} and tabulated in Tabs.~\ref{tab_uranus}--\ref{tab_ngc7027}. For Uranus and Neptune, we also include low resolution observations.  Note that for Neptune, Ceres and NGC~7027 we used a $\delta$-function for the source shape.

\begin{table*}[p]
\caption{Pointing offset results for Uranus. The derived median offset is calculated using frequencies from 1100 to 1400 GHz, the errors include the MAD error from the method and the $0.3\arcsec$ pointing jitter. All Uranus sparse observations centred on SSWD4 with LR and HR are included. Observations performed at two Beam-Steering Mirror position are labelled with ``bsm\_old'' and ``bsm\_new''. The bsm\_old position was used for all observations before \od{1011}, those after were taken with the BSM at the new home position bsm\_new. The difference between bsm\_old and bsm\_new is equivalent to $1.7\arcsec$ offset on the sky.}
\label{tab_uranus}
\begin{tabular}{rllll}
\hline\hline
OD & \multicolumn{1}{c}{OBSID (hex)} & Res & Offset ($\arcsec$) & Comments\\
\hline
1313 & \texttt{1342257307 (0x5001389B)} & HR & $  0.0 \pm   0.3$ & \textsc{<ref>} \\ 
 & \texttt{1342257305 (0x50013899)} & LR & $  0.7 \pm   0.5$ & \\ \hline
209 & \texttt{1342187880 (0x50002968)} & HR & $  1.6 \pm   0.3$ & \\ 
383 & \texttt{1342197472 (0x50004EE0)} & HR & $  0.6 \pm   0.3$ & \\ 
395 & \texttt{1342198273 (0x50005201)} & HR & $  0.4 \pm   0.3$ & \\ 
423 & \texttt{1342200175 (0x5000596F)} & HR & $  1.8 \pm   0.3$ & \\ 
571 & \texttt{1342210844 (0x5000831C)} & HR & $  2.7 \pm   0.3$ & \\ 
602 & \texttt{1342212338 (0x500088F2)} & HR & $  1.2 \pm   0.3$ & \\ 
767 & \texttt{1342222864 (0x5000B210)} & HR & $  0.4 \pm   0.3$ & \\ \hline
972 & \texttt{1342237013 (0x5000E955)} & HR & $  2.8 \pm   0.3$ & \\
 & \texttt{1342237014 (0x5000E956)} & HR & $  3.1 \pm   0.3$ & \\ 
 & \texttt{1342237016 (0x5000E958)} & LR & $  2.8 \pm   0.4$ & \\ 
 & \texttt{1342237017 (0x5000E959)} & HR & $  2.7 \pm   0.3$ & bsm\_old \\ 
 & \texttt{1342237017 (0x5000E959)} & HR & $  1.8 \pm   0.3$ & bsm\_new \\ \hline
1112 & \texttt{1342246283 (0x50010D8B)} & LR & $  4.2 \pm   0.3$ & bsm\_old \\ 
 & \texttt{1342246283 (0x50010D8B)} & LR & $  2.6 \pm   0.3$ & bsm\_new \\ 
 & \texttt{1342246285 (0x50010D8D)} & HR & $  3.6 \pm   0.3$ & bsm\_old \\ 
 & \texttt{1342246285 (0x50010D8D)} & HR & $  1.8 \pm   0.3$ & bsm\_new \\ \hline
1125 & \texttt{1342246974 (0x5001103E)} & HR & $  4.2 \pm   0.3$ & \\ 
 & \texttt{1342246975 (0x5001103F)} & LR & $  4.1 \pm   0.3$ & \\ \hline
1130 & \texttt{1342247100 (0x500110BC)} & HR & $  1.5 \pm   0.3$ & \\ 
 & \texttt{1342247101 (0x500110BD)} & LR & $  1.9 \pm   0.4$ & \\ \hline 
1145 & \texttt{1342247618 (0x500112C2)} & HR & $  2.0 \pm   0.3$ & \\ 
 & \texttt{1342247619 (0x500112C3)} & LR & $  2.1 \pm   0.3$ & \\  \hline
1151 & \texttt{1342247767 (0x50011357)} & HR & $  3.4 \pm   0.3$ & \\ 
 & \texttt{1342247766 (0x50011356)} & LR & $  3.5 \pm   0.3$ & \\  \hline
1326 & \texttt{1342257926 (0x50013B06)} & HR & $  1.5 \pm   0.3$ & \\ 
 & \texttt{1342257927 (0x50013B07)} & LR & $  1.4 \pm   0.4$ & \\  \hline
1335 & \texttt{1342258697 (0x50013E09)} & HR & $  1.5 \pm   0.3$ & \\ 
 & \texttt{1342258696 (0x50013E08)} & LR & $  1.6 \pm   0.4$ & \\  \hline
1341 & \texttt{1342259588 (0x50014184)} & HR & $  3.0 \pm   0.3$ & \\ 
 & \texttt{1342259587 (0x50014183)} & LR & $  3.5 \pm   0.3$ & \\  \hline
\end{tabular}
\end{table*}

\subsection{Uranus}

The results for Uranus, for the central detector SSWD4, are shown in Tab.~\ref{tab_uranus} and in Fig.~\ref{fig_results} (upper left panel). The current point source calibration in \textsc{spire\_cal\_11\_0} uses deep (22 repetitions) Uranus observations: \obsid{1342197472} (HR) from \od{383} and \obsid{1342237016} (LR) from \od{972} for observations before \od{1011} and \obsid{1342246285} and \obsid{1342246283} from \od{1112} for HR and LR respectively, for observations after \od{1011} (see \citealt{swinyard13} for more details). In \od{1112}, the observations of Uranus were followed by a specially designed LR map on the target, which was used to derive a pointing offset of $(2.0\pm0.2)\arcsec$. This offset was used to correct the ``after'' observations of Uranus before using them for point source conversion. Hence it is important to confirm  the derived correction by an independent estimation. The results of the method, shown in Tab.~\ref{tab_uranus}, for the two Uranus observations performed on the same \od{1112} are $(1.8\pm0.3)\arcsec$ (HR, bsm\_new case) and $(2.6\pm0.3)\arcsec$ (LR, bsm\_new case). The agreement is within the uncertainties for the HR observation, while it is marginally consistent with the LR observation. A similar LR map was taken for the observations of Uranus in \od{1313} and the map derived offset is $0.6\arcsec$. Our method reproduces this offset quite well: the LR observation \obsid{1342257305} is at $0.7\arcsec$, while our reference HR observation \obsid{1342257307} is by definition at zero offset. This is in support that Uranus during the reference observation was indeed well centred. This also provides a good consistency check for the method.

Additional consistency check of the method are the two special Uranus observations in \od{1112}: \obsid{1342246285} (HR) and \obsid{1342246283} (LR), with the beam-steering mirror at the two positions ``bsm\_old'' and ``bsm\_new''. From the results in Tab.~\ref{tab_uranus} we see a difference of 1.6\arcsec\ (LR) and 1.8\arcsec\ (HR), which is in very good agreement with the known 1.7\arcsec\ offset between the two BSM positions. 

As can be seen from Fig.~\ref{fig_results} and Tab.~\ref{tab_uranus}, there is no systematically worse pointing for Uranus observations before \od{1011}, when the BSM mirror home position was at 1.7\arcsec\ with respect to the optical axis. We attribute this to the small BSM offset, in comparison to the smallest FTS beam, as well as the similar magnitude of the APE uncertainty, which makes it such that no systematic effects are seen.

\subsection{Uranus on off-axis detectors}

In addition to the observations of Uranus in the central detectors (SSWD4 and SLWC3), we also performed observations on the unvignetted SSW off-axis detectors, i.e. those with centres within $\sim1^{\prime}$ from SSWD4. These observations were used to derive the point source conversion for the off-axis detectors. This calibration is useful when there are serendipitous point sources in the field of the FTS that fall on an off-axis detector or in the case of spectral mapping observations with both point sources and extended emission. As this calibration is also important, we assess the pointing accuracy of the associated Uranus observations.

The results after applying the relative pointing method, using as reference \obsid{1342257307} from \od{1313}, are shown in Tab.~\ref{tab_uranus2} and they indicate that the pointing offsets vary significantly: from 2-3\arcsec to 6-7\arcsec. These large pointing offsets reflect the fact that the accurate positions of the off-axis detectors, measured during the Herschel Performance Verification phase, were not updated in the Spacecraft-to-Instrument Alignment Matrix (SIAM) --- for some detectors the error is $\sim6\arcsec$. The SIAM was used to move the telescope in order to center the target on a particular off-axis detector for the observations in \od{767} and \od{972}, while a differential telescope offset, with respect to the central detector, was used during the observations in \od{383}. This can explain the systematically better pointing results for \od{383} observations. The reason that the SIAM apertures were never corrected for off-axis detectors was that all standard science observations were aligned using the SIAM for the central detectors (i.e. the error only affects this particular set of calibration observations).


With the proposed method and results we can properly calibrate the off-axis detectors with Uranus, after correcting them for the pointing offsets. This will improve significantly the point-source flux calibration, as in some cases the correction is up to 30-35\%. This has implications for the overall flux calibration of the off-axis detectors and also for mapping observations, when they are converted to point-source flux density units (Jy).

Regarding the off-axis detectors in SLW, some of them are co-aligned with SSW detectors, although the SLW beam centres do not perfectly match the SSW beam centres. This intrinsic misalignment is not taken into account in the current calibration scheme, however we expect the effects of this correction to be at less than a 3\% level, because of the large SLW beam.

\begin{table}[p]
\caption{Pointing offset results for Uranus on off-axis detectors. See Tab.~\ref{tab_uranus} for details.}
\label{tab_uranus2}
\begin{tabular}{lrll}
\hline\hline
Detector & OD & \multicolumn{1}{c}{OBSID (hex)} & Offset ($\arcsec$) \\
\hline
SSWC2/SLWD3 & 383 & \texttt{1342197477 (0x50004EE5)} & $3.0 \pm 0.6$ \\
SSWC5/SLWB3 & 383 & \texttt{1342197476 (0x50004EE4)} & $2.0 \pm 0.7$ \\
SSWE5/SLWB2 & 383 & \texttt{1342197473 (0x50004EE1)} & $2.9 \pm 0.5$ \\
SSWF3/SLWC2 & 383 & \texttt{1342197475 (0x50004EE3)} & $3.3 \pm 0.5$ \\
SSWB3/SLWC4 & 383 & \texttt{1342197478 (0x50004EE6)} & $2.3 \pm 0.6$ \\
SSWE2/SLWD2 & 383 & \texttt{1342197474 (0x50004EE2)} & $3.4 \pm 0.5$ \\ \hline
SSWE4 & 767 & \texttt{1342222865 (0x5000B211)} & $6.0 \pm 0.4$ \\
SSWE3 & 767 & \texttt{1342222866 (0x5000B212)} & $6.7 \pm 0.4$ \\
SSWD3 & 767 & \texttt{1342222867 (0x5000B213)} & $5.3 \pm 0.4$ \\
SSWC3 & 767 & \texttt{1342222868 (0x5000B214)} & $5.2 \pm 0.4$ \\
SSWC4 & 767 & \texttt{1342222869 (0x5000B215)} & $3.6 \pm 0.5$ \\ \hline
SSWD2 & 972 & \texttt{1342237021 (0x5000E95D)} & $7.1 \pm 0.4$ \\
SSWD6 & 972 & \texttt{1342237022 (0x5000E95E)} & $4.4 \pm 0.5$ \\
SSWB2 & 972 & \texttt{1342237019 (0x5000E95B)} & $2.8 \pm 0.6$ \\
SSWB4 & 972 & \texttt{1342237020 (0x5000E95C)} & $2.5 \pm 0.5$ \\
SSWF2 & 972 & \texttt{1342237023 (0x5000E95F)} & $6.1 \pm 0.4$ \\
\end{tabular}
\end{table}

\subsection{Neptune}

The relative pointings for Neptune are shown in Tab.~\ref{tab_neptune} and in Fig.~\ref{fig_results}, upper right panel. The results for the observations in \od{1112}, at the two BSM positions, are not showing the expected difference of $1.7\arcsec$. We attribute this discrepancy to the unfortunate direction on the sky of the 1.7$\arcsec$ BSM offset (which is always fixed in instrument coordinates) from the old to the new position, placing the planet between both beam centres.

Similarly to Uranus, we see a good agreement between observations performed during the same day and observations in LR and HR mode. There is no systematic off-pointing for observations before \od{1011}.

\begin{table*}[p]
\caption{Pointing offset results for Neptune. See Tab.~\ref{tab_uranus} for details.}
\label{tab_neptune}
\begin{tabular}{rllll}
\hline\hline
OD & \multicolumn{1}{c}{OBSID(hex)} & Res & Offset ($\arcsec$) \\
\hline
1080 & 1342245081 \texttt{(0x500108D9)} & HR & $  0.0 \pm   0.3$ & \\ 
 & 1342245080 \texttt{(0x500108D8)} & LR & $  0.5 \pm   0.4$ & \\ \hline
209 & 1342187887 \texttt{(0x5000296F)} & HR & $  0.6 \pm   0.4$ & \\ 
342 & 1342195348 \texttt{(0x50004694)} & HR & $  1.6 \pm   0.3$ & \\ 
354 & 1342195771 \texttt{(0x5000483B)} & HR & $  0.8 \pm   0.5$ & \\ 
368 & 1342196617 \texttt{(0x50004B89)} & HR & $  0.4 \pm   0.4$ & \\ 
382 & 1342197368 \texttt{(0x50004E78)} & HR & $  1.4 \pm   0.3$ & \\ 
392 & 1342198429 \texttt{(0x5000529D)} & HR & $  1.5 \pm   0.3$ & \\ 
543 & 1342208385 \texttt{(0x50007981)} & HR & $  2.2 \pm   0.3$ & \\ 
557 & 1342209855 \texttt{(0x50007F3F)} & HR & $  3.9 \pm   0.3$ & \\ 
571 & 1342210841 \texttt{(0x50008319)} & HR & $  1.4 \pm   0.3$ & \\ 
711 & 1342219564 \texttt{(0x5000A52C)} & HR & $  1.3 \pm   0.3$ & \\ 
742 & 1342221703 \texttt{(0x5000AD87)} & HR & $  1.2 \pm   0.4$ & \\ 
908 & 1342231992 \texttt{(0x5000D5B8)} & HR & $  3.1 \pm   0.3$ & \\ \hline
1099 & 1342245866 \texttt{(0x50010BEA)} & HR & $  0.7 \pm   0.4$ & \\ 
 & 1342245865 \texttt{(0x50010BE9)} & LR & $  1.2 \pm   0.4$ & \\ \hline
1112 & 1342246280 \texttt{(0x50010D88)} & HR & $  2.1 \pm   0.3$ & bsm\_old \\ 
 & 1342246280 \texttt{(0x50010D88)} & HR & $  1.8 \pm   0.3$ & bsm\_new \\ \hline
1125 & 1342246977 \texttt{(0x50011041)} & HR & $  0.3 \pm   0.4$ & \\ 
 & 1342246976 \texttt{(0x50011040)} & LR & $  0.8 \pm   0.4$ & \\ \hline
1284 & 1342255278 \texttt{(0x500130AE)} & HR & $  1.6 \pm   0.4$ & \\  
 & 1342255277 \texttt{(0x500130AD)} & LR & $  1.6 \pm   0.4$ & \\ \hline
1292 & 1342256096 \texttt{(0x500133E0)} & HR & $  1.9 \pm   0.4$ & \\ 
  & 1342256095 \texttt{(0x500133DF)} & LR & $  2.1 \pm   0.3$ & \\ \hline
1314 & 1342257352 \texttt{(0x500138C8)} & HR & $  1.4 \pm   0.3$ & \\ 
 & 1342257350 \texttt{(0x500138C6)} & LR & $  1.4 \pm   0.4$ & \\  \hline
\end{tabular}
\end{table*}

\begin{table}[p]
\caption{Pointing offset results for Ceres. See Tab.~\ref{tab_uranus} for details.}
\label{tab_ceres}
\begin{tabular}{rlll}
\hline\hline
OD & \multicolumn{1}{c}{OBSID (hex)} & Offset ($\arcsec$) \\
\hline
1187 & 1342249467 \texttt{(0x500119FB)} & $  0.0 \pm   0.3$  \\ 
275 & 1342190673 \texttt{(0x50003451)} & $  2.3 \pm   0.4$  \\ 
288 & 1342191220 \texttt{(0x50003674)} & $  3.5 \pm   0.4$  \\ 
327 & 1342193667 \texttt{(0x50004003)} & $  5.4 \pm   0.4$  \\ 
494 & 1342204878 \texttt{(0x50006BCE)} & $  3.5 \pm   0.4$  \\ 
767 & 1342222862 \texttt{(0x5000B20E)} & $  1.6 \pm   0.5$  \\ 
972 & 1342237010 \texttt{(0x5000E952)} & $  3.1 \pm   0.4$  \\ 
\hline
\end{tabular}
\end{table}

\begin{table}[p]
\caption{Pointing offset results for NGC\,7027, assuming a $\delta$-function for the source. Using only the SSW continuum, after a simultaneous fit to the lines and a second degree polynomial continuum. Only observations with HR are used. See Tab.~\ref{tab_uranus} for details. The polynomial fit for \obsid{1342195347} from \od{342} (italisized) overestimated the true continuum level and was performed by masking the noisier parts in the NGC\,7027 spectrum.}
\label{tab_ngc7027}
\begin{tabular}{rll}
\hline\hline
OD & \multicolumn{1}{c}{OBSID (hex)} & Offset ($\arcsec$) \\
\hline
1125 & 1342246971 \texttt{(0x5001103B)} & $  0.0 \pm   0.0$ \\ 
217 & 1342188197 \texttt{(0x50002AA5)} & $  2.7 \pm   0.3$ \\ 
227 & 1342188670 \texttt{(0x50002C7E)} & $  2.6 \pm   0.3$ \\ \hline
240 & 1342189121 \texttt{(0x50002E41)} & $  1.4 \pm   0.3$ \\ 
240 & 1342189124 \texttt{(0x50002E44)} & $  2.0 \pm   0.3$ \\ 
240 & 1342189125 \texttt{(0x50002E45)} & $  2.1 \pm   0.3$ \\ \hline
250 & 1342189543 \texttt{(0x50002FE7)} & $  3.1 \pm   0.3$ \\ 
326 & 1342193812 \texttt{(0x50004094)} & $  5.9 \pm   0.3$ \\ 
\textit{342} & \textit{1342195347 \texttt{(0x50004693)}} &  $\mathit{1.1 \pm   0.7}$ \\ 
368 & 1342196614 \texttt{(0x50004B86)} & $  1.6 \pm   0.3$ \\ 
383 & 1342197486 \texttt{(0x50004EEE)} & $  1.3 \pm   0.3$ \\ 
404 & 1342198921 \texttt{(0x50005489)} & $  1.2 \pm   0.3$ \\ 
557 & 1342209856 \texttt{(0x50007F40)} & $  2.5 \pm   0.3$ \\ 
572 & 1342210858 \texttt{(0x5000832A)} & $  2.7 \pm   0.3$ \\ 
601 & 1342212324 \texttt{(0x500088E4)} & $  1.3 \pm   0.3$ \\ 
711 & 1342219571 \texttt{(0x5000A533)} & $  0.6 \pm   0.4$ \\ 
742 & 1342221697 \texttt{(0x5000AD81)} & $  1.1 \pm   0.3$ \\ 
908 & 1342231993 \texttt{(0x5000D5B9)} & $  4.1 \pm   0.3$ \\ 
971 & 1342237007 \texttt{(0x5000E94F)} & $  3.4 \pm   0.3$ \\ 
988 & 1342238245 \texttt{(0x5000EE25)} & $  3.3 \pm   0.3$ \\ 
1053 & 1342243593 \texttt{(0x50010309)} & $  1.3 \pm   0.3$ \\ 
1080 & 1342245075 \texttt{(0x500108D3)} & $  1.1 \pm   0.3$ \\ 
1099 & 1342245861 \texttt{(0x50010BE5)} & $  0.4 \pm   0.4$ \\ 
1111 & 1342246255 \texttt{(0x50010D6F)} & $  0.7 \pm   0.3$ \\ 
1130 & 1342247106 \texttt{(0x500110C2)} & $  1.1 \pm   0.3$ \\ 
1145 & 1342247623 \texttt{(0x500112C7)} & $  1.0 \pm   0.3$ \\ 
1283 & 1342255260 \texttt{(0x5001309C)} & $  2.4 \pm   0.3$ \\ 
1313 & 1342257340 \texttt{(0x500138BC)} & $  2.5 \pm   0.3$ \\ 
1325 & 1342257918 \texttt{(0x50013AFE)} & $  1.8 \pm   0.3$ \\ 
1341 & 1342259592 \texttt{(0x50014188)} & $  3.0 \pm   0.3$ \\ 
\hline
\end{tabular}
\end{table}

\subsection{Ceres}

Ceres is another SSO included in the analysis, with 7 observations all in HR mode, for which the results are show in Tab.~\ref{tab_ceres} and in Fig.~\ref{fig_results}, bottom left. The results can be used to improve the current Ceres models. There are other asteroids with repeated FTS observations and this method can be applied to correct for pointing effects in their scientific analysis (Lim et al, in preparation).

\subsection{NGC\,7027}

The results for NGC\,7027 are shown in Tab.~\ref{tab_ngc7027} and in Fig.~\ref{fig_results}, bottom right. The case of NGC\,7027 is sightly complicated because the source has many emission lines and in order to apply the reference ratio method we need to extract the continuum. We did this by simultaneously fitting  the brightest lines and the continuum in the level-2 spectrum. The fitted continuum --- a polynomial of a second order --- is then used to calculate $R_{obs}$.  We note that the use of a second order polynomial is potentially a poor fit at frequencies above 1400 GHz and below 1100 GHz (see Fig.~\ref{fig_examples}), but as we limit the median estimation within [1110,1400] GHz then the effects do not bias the results. There was only one peculiar case (in italic in Tab.~\ref{tab_ngc7027}) where we had to explicitly mask the spectra near the band edges, before performing the polynomial fit. Without this masking the fit was poor and the polynomial fit was significantly overestimating the continuum.

In \od{326} \textit{Herschel} suffered a general telescope mispointing, due to the use of a wrong SIAM. The only observation we have in this OD is for NGC\,7027 (\obsid{1342193812}) and the derived pointing offset of $5.9\arcsec$. This is in excellent agreement with the known offset of the SIAM on \od{326} ($\sim6\arcsec$), giving another good consistency check of the method.

\subsection{Illustration of the pointing correction}

\begin{figure}
\centerline{
\includegraphics[width=0.5\textwidth]{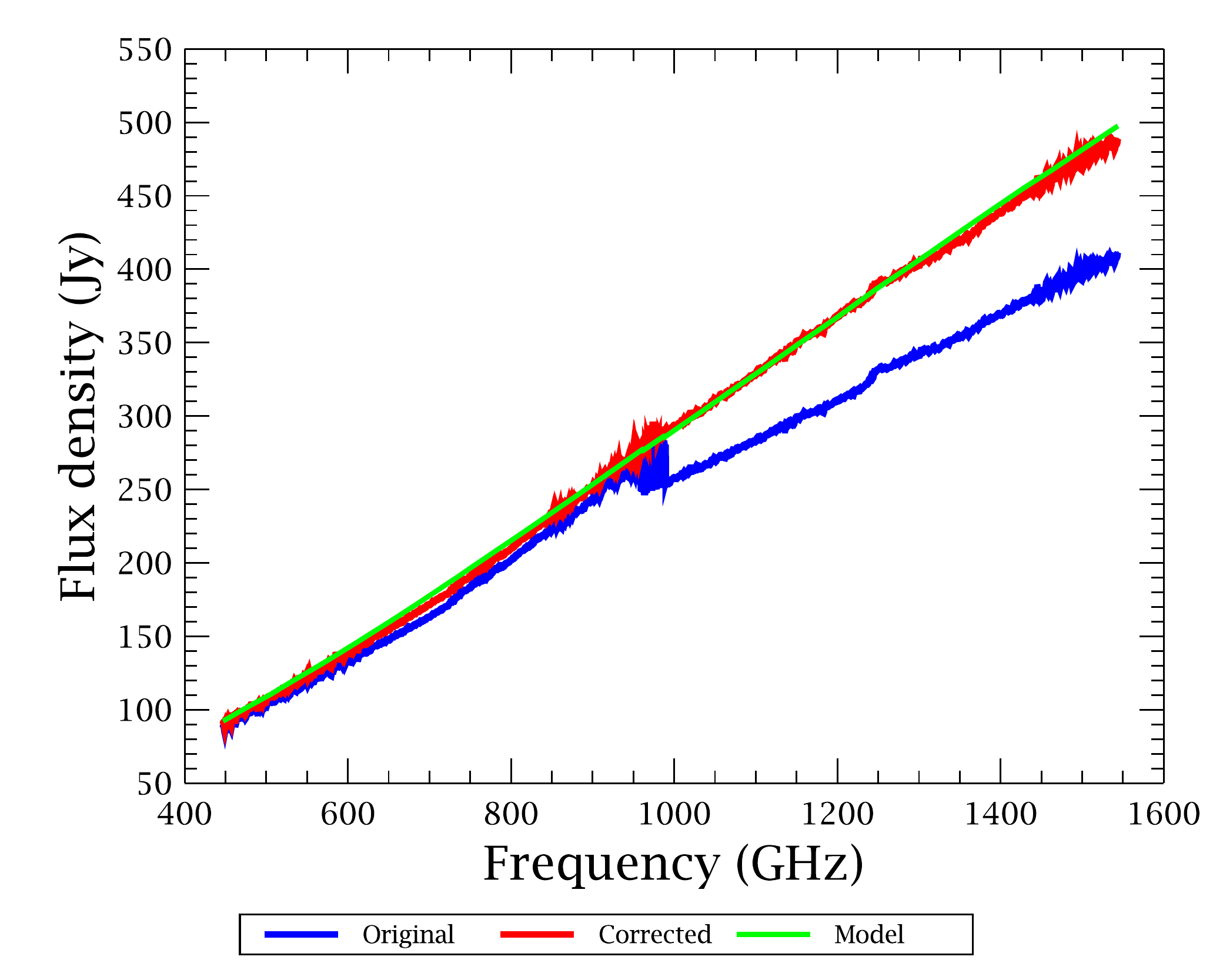}}
\caption{Applying the derived pointing offset to one Uranus observation: \obsid{1342246974} from \od{1125}. The original SSW and SLW spectra from the central detectors are shown in blue. The corrected spectra, taking the derived offset of 4.2\arcsec, are shown in red. The Uranus predicted spectrum for the time of the observation is shown in green.}
\label{fig_out}
\end{figure}

To illustrate how the results of this study can be used to correct spectra for pointing offsets, and consequently to improve the FTS calibration, we selected one Uranus observation, \obsid{1342246974} from \od{1125} -- one of the most deviating ones, with a derived pointing offset of 4.2\arcsec. Fig.~\ref{fig_out} shows the original spectra from the central detectors SSWC3 and SSWD4 (in blue) and the corrected spectra, assuming a pointing offset of 4.2\arcsec (in red). While the correction for the SLW spectrum is less than 5\%, we see a dramatic change in the continuum level of SSW, which improves significantly the spectral shape and the stitching of the two FTS bands. The corrected spectrum is also much closer to the model prediction (shown in green), which gives a good consistency check for the method. This example shows the potential of the method for improvements of the spectra of calibration targets and the overall calibration.

The illustration shown in Fig.~\ref{fig_out} gives an interesting idea that, even in the case of not knowing the pointing of a given observation of a \textit{point-source}, the difference between SSW and SLW in the overlap region could be used as a crude guess on the pointing offset. Of course this can only be done for point sources with high signal-to-noise, because this overlap region is the noisiest part of the FTS spectra. Nevertheless, this simple alternative way of correcting the effect of pointing is worth considering as it could be applied to science observations which do not have a well pointed reference observation for comparison.

\section{Conclusions and further prospects}
\label{sec:conclusions}

The method presented in this paper provides means to derive the relative pointing offsets for FTS data, with respect to a reference observation. It is important to note that with this method we cannot retrieve the actual pointing, but only the offset relative to a selected observation. The greater the number of observations of a particular calibration target the higher the chances there will be an observation or observations in which the target will be well centred in the beam.

Our conclusions are mainly focused on the results for Uranus, as this planet is used as the primary FTS calibrator. The results on the other calibration targets are also valuable and can be used in a number of studies that may improve our understanding of the accuracy and the repeatability of the line and continuum measurements.


The derived relative pointing offsets for Uranus unequivocally indicate that during \obsid{1342257307} from \od{1313}, our reference observation, the planet was very close to the centre of the SSW beam. This was also confirmed by the low resolution spectral map of Uranus taken directly after the sparse observation, with inferred position of $0.6\arcsec$ of Uranus with respect to the expected position. This is the uncertainty in the transition from relative to absolute pointing offsets for Uranus. In addition, we presented a number of consistency checks in support of the method and we are confident that the reported pointing offsets are correct, within the uncertainties introduced by the telescope jitter and the planets models. 

Although \obsid{1342257307} from \od{1313} is better centred in the beam and no pointing correction is needed, it is significantly shallower, with only 4 repetitions, in comparison with the current observations used for the HR mode point-source calibration: \obsid{1342197472} from \od{383} and \obsid{1342246285} from \od{1112}, both with 22 repetitions.  Hence, it is not advisable to use it for the point source conversion because of the lower signal to noise in the Uranus spectrum.

We confirm the pointing offsets of the deep Uranus observations from \od{1112} used in the current flux calibration scheme. We note that the flux calibration used up to HIPE v10, for all observations, was based on the deep observation \obsid{1342197472} from \od{383}. Our results clearly indicated that Uranus was very close to the beam centre, at a sub-arcsec offset, during this observation. This means the current FTS point-source flux calibration, used for observations taken before the BSM change (or for all observation before HIPE v10), is of a similar accuracy to the calibration now adopted for observations after the BSM change, which uses an observation corrected for pointing.

The results we provide for Uranus observations on a number of off-axis detectors indicate up to 6-7\arcsec pointing offsets. These were not taken into account in the current (\textsc{SPIRE\_CAL\_11\_0}) flux calibration, which means that there is a large overestimation of the conversion factors. Therefore, the interpretation of point-source calibrated data from the off-axis detectors or from spectral maps must be carefully reconsidered for a possible flux loss. The results from this study will be incorporated in the next update of the FTS calibration.

Having the correct pointing offsets makes it possible to combine all available observations of a given calibration target, because we can effectively correct each one for the pointing. This procedure will decrease the overall noise in the spectrum,  however, one has to be careful because the systematic noise properties of observations at different off-axis positions will be different. Nevertheless we envisage to apply this idea to the upcoming calibration of the FTS and evaluate the pros and cons with respect to the current calibration scheme. 

The improvements in the FTS calibration, incorporating the results from this study, are part of the on-going work of the FTS team.

\section*{Acknowledgments}
SPIRE has been developed by a consortium of institutes led by Cardiff University (UK) and including Univ. Lethbridge (Canada); NAOC (China); CEA, LAM (France); IFSI, Univ. Padua (Italy); IAC (Spain); Stockholm Observatory (Sweden); Imperial College London, RAL, UCL-MSSL, UKATC, Univ. Sussex (UK); and Caltech, JPL, NHSC, Univ. Colorado (USA). This development has been supported by national funding agencies: CSA (Canada); NAOC (China); CEA, CNES, CNRS (France); ASI (Italy); MCINN (Spain); SNSB (Sweden); STFC (UK); and NASA (USA).

\end{document}